\documentclass{article}

\usepackage{graphicx}
\usepackage{amssymb,amsmath,amsfonts}
\usepackage{bm}
\usepackage{subfigure}
\usepackage{graphicx}
\usepackage{epstopdf}

\usepackage{natbib}
\usepackage{url}
\usepackage[skip=2pt,font=small]{caption}

\usepackage[a4paper,left=1in,top=1in,right=1in,bottom=1in,ignoreheadfoot]{geometry}
\usepackage{algorithm}

\usepackage{threeparttable}
\usepackage{pgf}
\usepackage{tikz}
\usepackage{subfig}
\usepackage[toc,page]{appendix}
\usepackage{authblk}
\usetikzlibrary{arrows,automata}

\graphicspath{{.}}

\DeclareMathOperator*{\Exp}{Exp}
\DeclareMathOperator*{\Gam}{Gamma}

\DeclareMathOperator{\Unif}{Unif}

\DeclareMathOperator{\DP}{DP}
\begin{document}

\title{Scalable Bayesian nonparametric regression via a Plackett-Luce model for conditional ranks}
\author{Tristan Gray-Davies\thanks{Email: tgdavies@stats.ox.ac.uk}, Chris C. Holmes\thanks{Email: cholmes@stats.ox.ac.uk}, Fran\c cois Caron\thanks{Email: caron@stats.ox.ac.uk}}
\affil{Department of Statistics, University of Oxford}

\maketitle


%



\begin{abstract}
We present a novel Bayesian nonparametric regression model for covariates $X$ and continuous response  variable $Y\in\mathbb R$. The model is parametrized in terms of marginal distributions for $Y$ and $X$ and a regression function which tunes the stochastic ordering of the conditional distributions $F(y|x)$. By adopting an approximate composite likelihood approach, we show that the resulting posterior inference can be decoupled for the separate components of the model.  This procedure can scale  to very large datasets and allows for the use of standard, existing,
software from Bayesian nonparametric density estimation and Plackett-Luce ranking estimation to be
applied. As an illustration, we show an application of our approach to a US Census dataset, with over
1,300,000 data points and more than 100 covariates.
\end{abstract}





%

\section{Introduction}
\label{sec:intro}

Bayesian nonparametric regression offers a flexible and robust way of modeling the dependence between covariates $x\in \mathcal X$ and a response variable $Y\in \mathbb R$ by using models with larger support than their parametric counterparts. Nonparametric statistical models are motivated by robustness and their ability to capture effects such as outliers, strong nonlinearities or multimodalities, while providing probabilistic measures of predictive uncertainty. Bayesian nonparametric regression methods are largely underpinned by one of two random probability measures namely, Dirichlet process mixtures~\citep{Ferguson1973,Lo1984} and P\'olya trees~\citep{Lavine1992,Lavine1994}. These approaches, widely applied to density estimation problems \citep[see e.g.][]{Hjort2010}, have been used as building blocks of various nonparametric regression models through a number of different approaches.

One approach, called the conditional approach, considers the covariates as fixed, and models directly the conditional distribution $f(y|x)$ of the response given the covariate . This conditional distribution may be constructed in a semiparametric or fully nonparametric way. The semiparametric conditional approach typically assumes that
\begin{align}
Y=\eta(x)+\epsilon
\end{align}
where $\eta$ is some unknown flexible mean function and $\epsilon$ is the residual. Regression models (priors) have been proposed for the mean function $\eta$ such as Gaussian processes \citep[see e.g.][]{Rasmussen2006}, basis function representations such as splines or kernels~\citep{Denison2002,Muller2004} or Bayesian regression trees~\citep{Chipman2010}. More generally, \cite{Kottas2001} and \cite{Lavine1995} proposed to use Dirichlet process mixtures for the distribution of the residuals, while \cite{Pati2014} jointly model the mean function and residual distribution using Gaussian processes and probit stick-breaking processes~\citep{Chung2009}.
The fully nonparametric conditional approach considers that $f(y|x)=\int_\Theta f(y|x,\theta)P_x(d\theta)$ takes the form of a mixture model with unknown mixing distribution $P_x$ for $\theta$. A prior is set on the family of probability distributions $(P_x)_{x\in\mathcal X}$. In particular, following the seminal work of \citet{MacEachern1999}, various dependent Dirichlet process models have been proposed in the literature~\citep{Gelfand2003,GriffinSteel2006,DunsonPillaiPark2007,Caron2007,Caron2008,Dunson2008}. Similarly, \cite{Trippa2011} define a class of dependent random probability distributions using P\'olya trees.

An alternative to the conditional approach is to treat the covariates as random variables and to build a joint statistical model for $(X,Y)$. In this way, one can cast the regression problem as a density estimation one. For example, \cite{Mueller96} proposed to use Dirichlet process mixtures for the joint distribution of $(X,Y)$. This approach was later extended by~\cite{Shahbaba2009}, \cite{Hannah2011} and \cite{Wade2014}.

A major drawback of current Bayesian methods for semi or nonparametric regression is that many methods do not scale well with the number of samples and/or with the dimensionality of the covariates. In this paper, we propose a novel joint Bayesian nonparametric regression model $F_{X,Y}$ that affords an approximation which can scale easily to large data applications. The model is parameterized in terms of the marginal distributions of the response $F_Y$ and covariates $F_X$, and then a conditional regression model that utilises the two marginal distributions,
\begin{align}
F_X &\sim \mathbb P_X\\
F_Y &\sim \mathbb P_Y\\
\beta&\sim \pi_\beta\\
F_{X,Y}(x,y)&=C_{\lambda_{\beta}}(F_X(x),F_Y(y))
\end{align}
where $\mathbb P_X$ and $\mathbb P_Y$ are some nonparametric prior over probability distributions, $\lambda_{\beta}:\mathcal X\rightarrow \mathbb R_+$ is some parametric regression function of the covariates, and $C_{\lambda_{\beta}}$ plays a role similar to a copula in that it takes marginal distributions as inputs and characterises the dependence between them using the function $\lambda_{\beta}$. In particular we consider a Plackett-Luce model for ranks for the regression structure. This construction, detailed in Section~\ref{sec:model}, builds on the original Plackett-Luce model~\citep{luce1959,plackett1975} for ranking. The positive function $\lambda_\beta$ tunes the stochastic ordering of the responses given the covariates, the ratio  $\lambda_\beta(X_i)/\left (\lambda_\beta(X_i)+\lambda_\beta(X_j)\right )$ representing the conditional probability, $Pr(Y_i < Y_j | X_i, X_j)$, that response $Y_i$ is less than response $Y_j$ given knowledge of $\{X_i, X_j\}$. There is thus a natural interpretation of the parameters: $\lambda_\beta$ tunes the relative ordering of the responses at different covariate values, and $F_Y$ sets the marginal distribution of the responses. This  strong interpretability is an important feature as it provides a good vehicle for specifying prior beliefs.

For inference we propose to use a marginal composite  likelihood approach, which we show allows the model to scale tractably to  large data applications and allows for the use of standard, existing, software from Bayesian nonparametric density estimation and Plackett-Luce ranking estimation to be applied. As an illustration, we show an application of our approach to a US Census dataset, with over 1,300,000 data points and more than 100 covariates.

The paper is organized as follows. Section~\ref{sec:densityestimation} provides background on Dirichlet process mixtures and P\'olya trees for density estimation. Section~\ref{sec:model} describes the Plackett-Luce copula model. The marginal composite likelihood approach for scalable inference is presented in Section~\ref{sec:inference}. Section~\ref{sec:appli} presents some results of our approach on simulated data and on the US Census dataset.

\section{Bayesian nonparametric density estimation}
\label{sec:densityestimation}
%

The appeal of Bayesian nonparametric models is the large support and probabilistic inference provided by such priors. This both safeguards against model misspecification and enables highly flexible estimation of distributions. This has lead to particular popularity of Bayesian nonparametric priors in density estimation.\smallskip


In the simple case of density estimation for a real valued random variable many nonparametric priors exist - see \cite{Hjort2010} for a recent review. A popular class of model is the Dirichlet Process Mixture (\cite{Lo1984}), whereby a Dirichlet process prior is placed on the distribution of the parameters of a parametric family. The result is an ``infinite mixture model''. Precisely:

\begin{align*}
f_Y(y)&= \int K(y | \theta) dP(\theta) \\
P &\sim \DP(c, P_0)
\end{align*}
where $K$ is the density of the chosen parametric family, $c>0$ is a scale parameter and $P_0$ is a base measure. Since draws from a Dirichlet Process are almost surely atomic measures, there is positive probability of observations sharing a parameter value given the random measure $P$. The result is an effect of clustering within a sample, with a random, limitless number of clusters. This has proved to be an extremely popular model as it models heterogeneity within a sample well, and provides a highly flexible support. Efficient MCMC schemes (\cite{Escobar1995,MaceachernMuller1998,neal2000}) have lead to the widespread use of the Dirichlet Process Mixture (DPM) in density estimation.


P\'olya trees provide another flexible nonparametric prior for density estimation (\cite{Ferguson1974,Lavine1992,Lavine1994,Mauldin1992}). They are defined as follows: Let $\bm{\epsilon} = (\epsilon_1, \ldots, \epsilon_k) \in E^k = \{ 0,1\}^k$, and define a sequence of embedded partitions of $\mathbb{R}$ to be $\Gamma_k = \{ B_{\bm{\epsilon}} : \bm{\epsilon} \in E^k\}$, where the $B_{\bm{\epsilon}}$ are defined recursively, such that $B_{\bm{\epsilon} 0} \cup B_{\bm{\epsilon} 1} = B_{\bm{\epsilon}}$. Now let $E^* = \cup_{k\geq 1}E^k$, the set of all countable sequences of zeros and ones, and let $\mathcal{A} = \{ \alpha_{\bm{\epsilon}} : \bm{\epsilon} \in E^*\}$ be a set of nonnegative real numbers. Then, a random probability measure $P$ is a P\'olya tree process with respect to $\Gamma = \{ \Gamma_k : k\geq1\}$ and $\mathcal{A}$ if $P( B_{\bm{\epsilon}0} \mid  B_{\bm{\epsilon}}) \sim \text{Beta}(\alpha_{\bm{\epsilon}0}, \alpha_{\bm{\epsilon}1})$, independently for all $\bm{\epsilon} \in E^*$. There are two properties of the P\'olya tree process that are appealing for density estimation: P\'olya trees are conjugate, meaning that both the prior and the posterior have the same functional form, and, for certain choices of $\mathcal A$, realizations are absolutely continuous probability distributions, almost surely. It is worth pointing out that empirically the model can depend heavily on the defined sequence of partitions $\Gamma$, although a mixture of P\'olya trees proposed by \cite{Lavine1992} can smooth out this dependence over multiple partitions. In what follows  we make use of these nonparametric models to specify priors for the marginal distributions of covariates and response variables.

\section{The statistical model}
\label{sec:model}

Let $(X_i,Y_i)$, $i=1,\ldots,n$ be the covariates and responses and regression function $\lambda_\beta:\mathcal X\rightarrow \mathbb R_+$. To build the dependence we introduce a latent random variable $Z_i$ that is used to capture the underlying relative level of the response via,
\begin{align}
Z_i|X_i=x_i\sim \Exp(\lambda_\beta(x_i))
\end{align}
 where $\Exp(a)$ denotes the standard exponential distribution of rate $a$. The latent variable $Z_i$ may be interpreted as an ``arrival time'' of individual $i$. The arrival times then define a conditional ranking of the predicted response variables $Y_1, \ldots, Y_n$.

The model can be summarized as follows, for $i=1,\ldots,n$
\begin{align}
X_i &\overset{\text{iid}}{\sim} F_X\\
Z_i \mid X_i,\beta &\overset{\text{ind}}{\sim} \Exp(\lambda_{\beta}(X_i))\\
Y_i &= F_Y^{-1}(F_Z(Z_i))
\end{align}

where
\begin{align*}
F_Z(z) &= \int_{\mathcal X} F_{Z \mid X=x}(z) dF_X(x)\\
& =  \int_{\mathcal X} \left (1 - e^{-\lambda_\beta(x)z}\right ) dF_X(x).
\end{align*}

\begin{figure}[h]
\begin{center}
\subfigure[]{\includegraphics[width=0.49\textwidth]{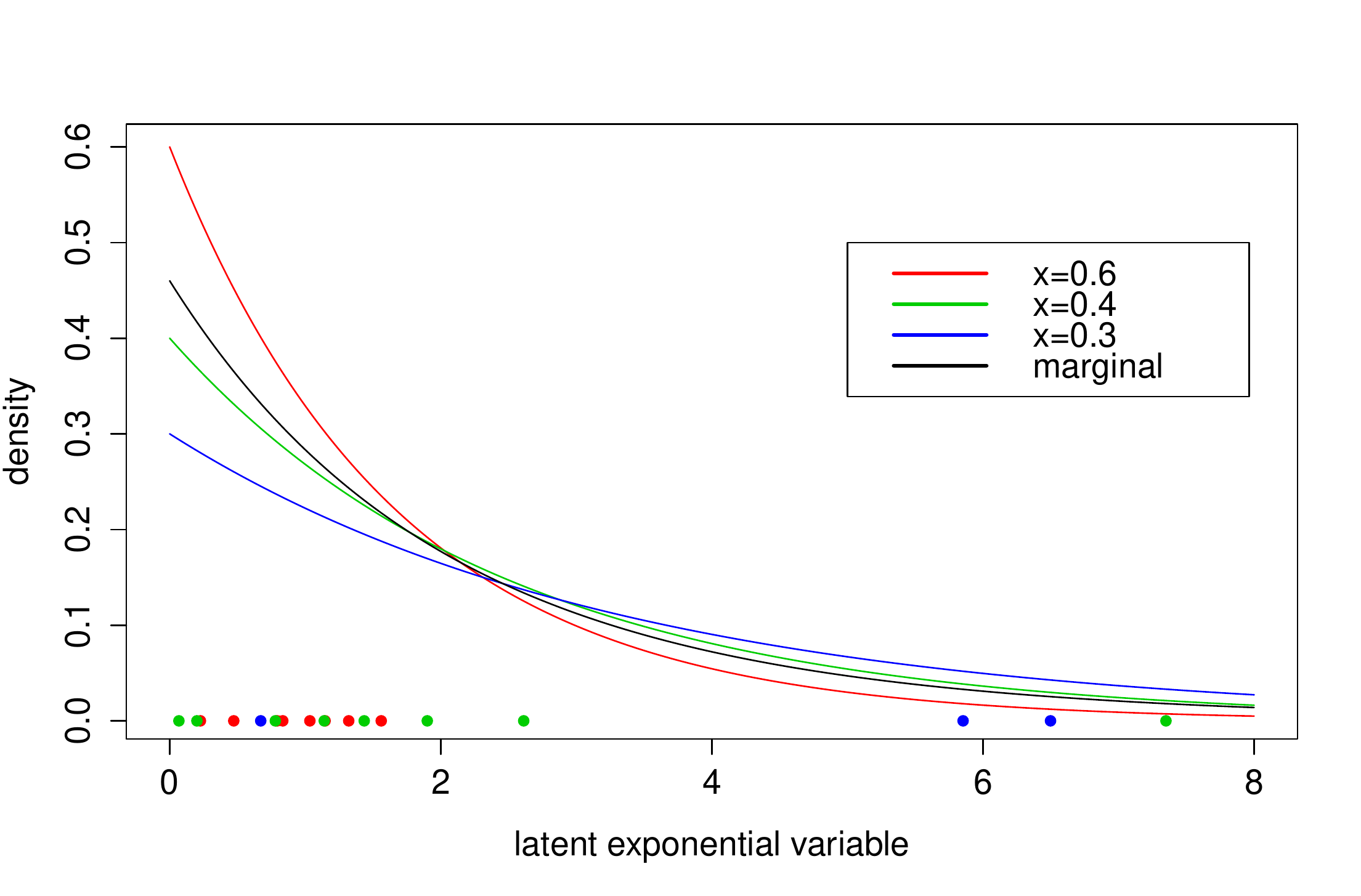}}
\subfigure[]{\includegraphics[width=0.49\textwidth]{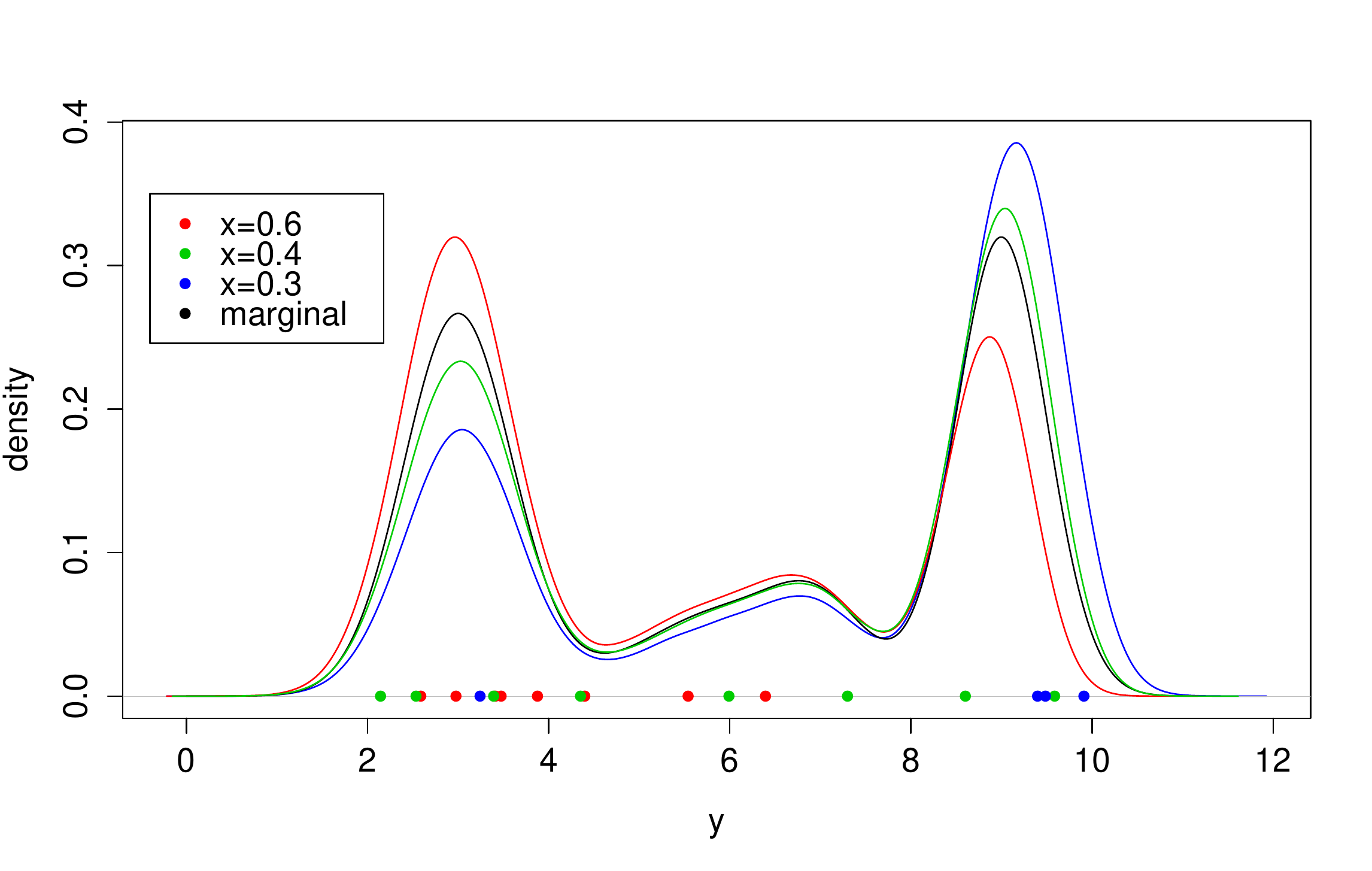}}
\end{center}	
\caption{Illustration of the latent variable used to capture the regression dependence. In (a) we show the distribution of the conditional latent variable, $Z$, at various points in $X$ assuming a log-linear dependence. In (b) we see the corresponding predictive distributions using a Gaussian mixture model for the marginal, $F_Y$, shown as the black line. The points in $Z$ shown in (a) are mapped to the points in $Y$ shown in (b) where the ordering is preserved.}
\label{fig:tree}
\end{figure}

Figure (\ref{fig:tree}) shows the correspondence between the conditional exponential random variables, $Z | X$, shown in \ref{fig:tree}(a) for differing covariate values, and the resulting predictive distributions in \ref{fig:tree}(b), where the marginal $F_Y$ is a Gaussian mixture model shown as the black line. We can see visually that the distributions in \ref{fig:tree}(b) are stochastically ordered under the model. The coloured points shown in (a) are mapped to the points shown in (b), where again ordering is preserved.

As $F_Y$ and $F_Z$ are cumulative density functions, $F_Y^{-1}\circ F_Z$ is a monotonically increasing function and
$$\mathbb{P}(Y_i \leq Y_j)= \mathbb{P}(Z_i \leq Z_j) = \frac{\lambda_\beta(x_i)}{\lambda_\beta(x_i)+\lambda_\beta(x_j)}.$$
This clarifies the role of the regression function.
More generally, given an ordering $\nu=(\nu_1,\ldots,\nu_n)$ (a permutation of $\{1, 2, \ldots, n\}$), we have

$$\mathbb{P}(Y_{\nu_1} \leq Y_{\nu_2}, \ldots, \leq Y_{\nu_n})= \mathbb{P}(Z_{\nu_1} \leq Z_{\nu_2}, \ldots, \leq Z_{\nu_n}) = \prod_{i=1}^n \frac{\lambda_\beta(x_{\nu_i})}{\sum_{j\geq i} \lambda_\beta(x_{\nu_j})}.$$

The above model is the Plackett-Luce model~\citep{luce1959,plackett1975}, popular in the ranking literature, and also corresponds to the partial likelihood used for Cox proportional hazards models \citep{Cox1972}.\medskip

By construction  $F_Z(Z_i)$ is marginally uniformly distributed on $[0,1]$. Thus, $Y_i = F_Y^{-1}( F_Z(Z_i))$ is marginally distributed from $F_Y$. The joint distribution $F_{X,Y}$ can thus be described in terms of marginals $F_X$ and $F_Y$ and a Plackett-Luce copula $C_{\lambda_{\beta}}$ such that
$$F_{X,Y}(x,y)=C_{\lambda_{\beta}}(F_X(x),F_Y(y)).$$

\begin{figure}[h]
\subfigure[$\lambda_\beta(x)=1$]{\includegraphics[width=.24\textwidth]{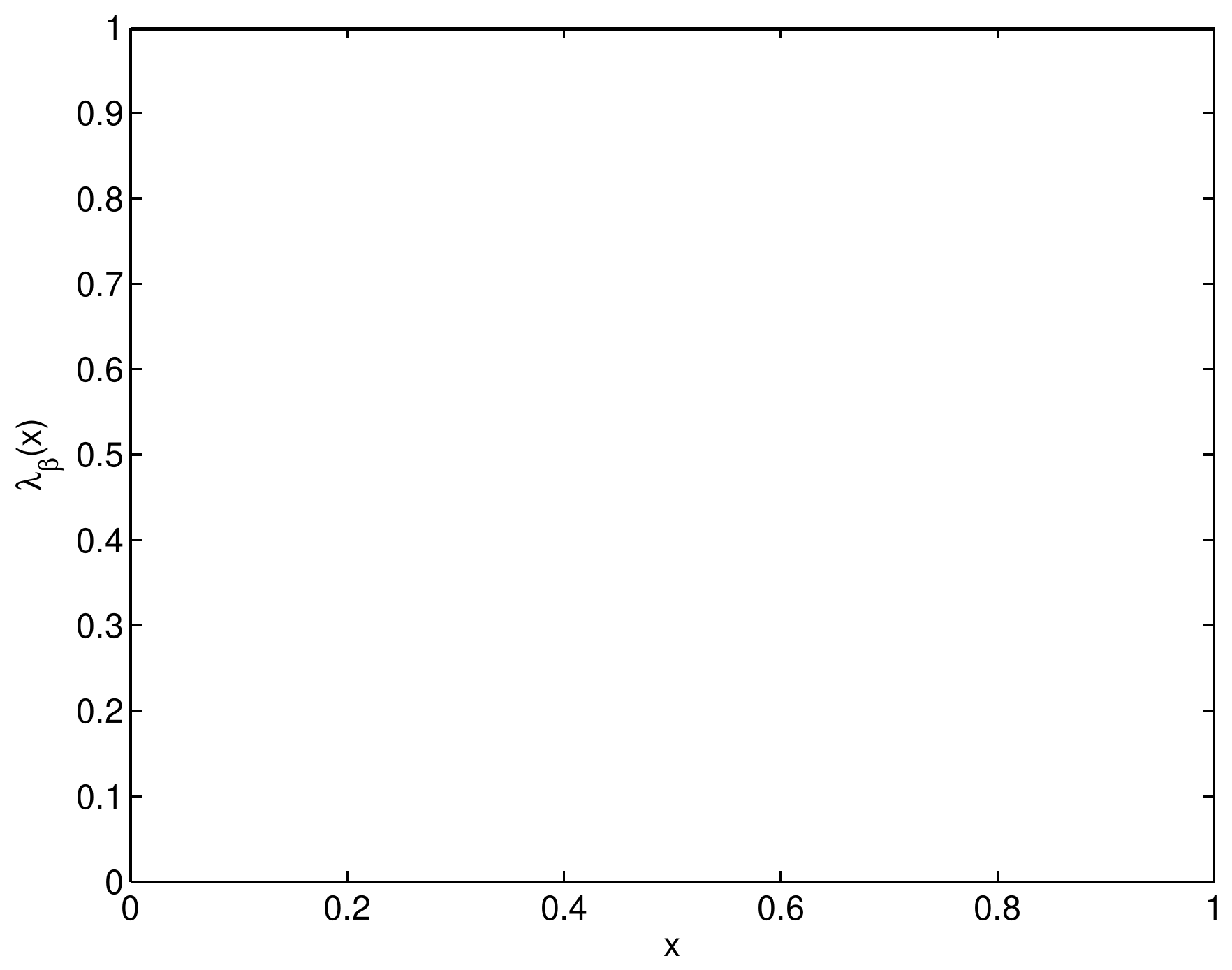}}
\subfigure[$\lambda_\beta(x)=.01$ if $x<0.5$, $1$ otherwise]{\includegraphics[width=.24\textwidth]{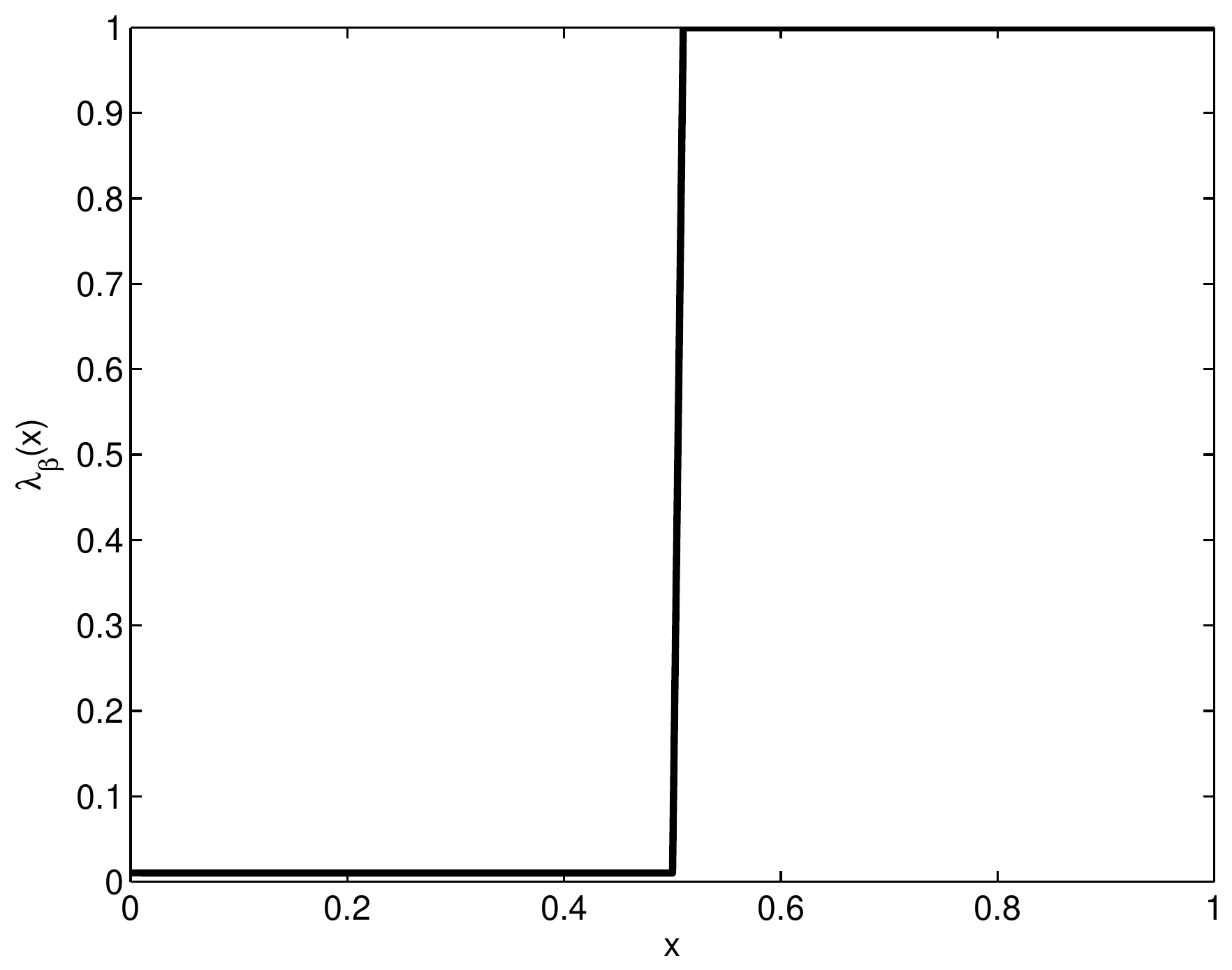}}
\subfigure[$\lambda_\beta(x)=\exp(-100(x-\frac{1}{2})^2)$]{\includegraphics[width=.24\textwidth]{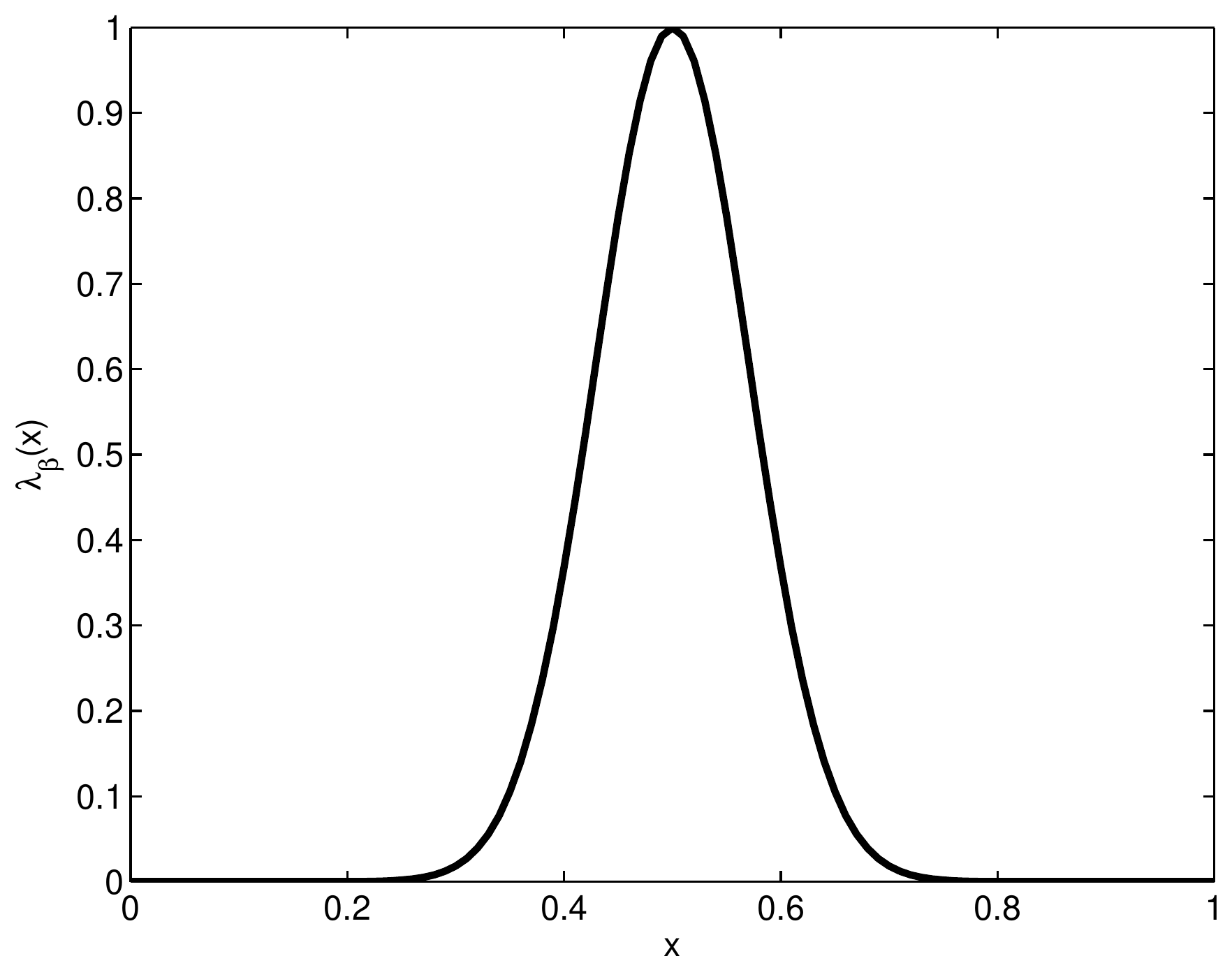}}
\subfigure[$\lambda_\beta(x)=\exp(-100x)$]{\includegraphics[width=.24\textwidth]{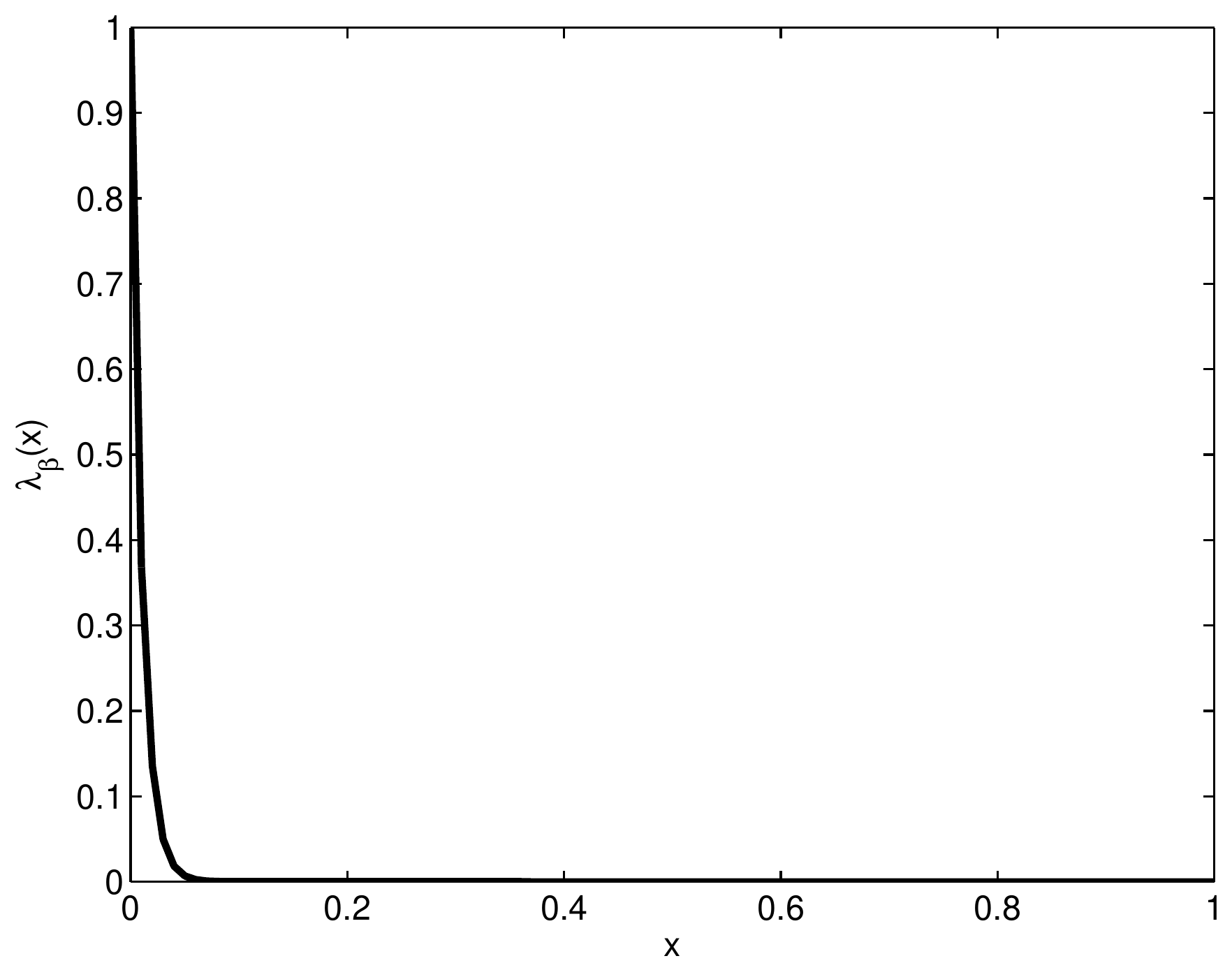}}
\subfigure[]{\includegraphics[width=.24\textwidth]{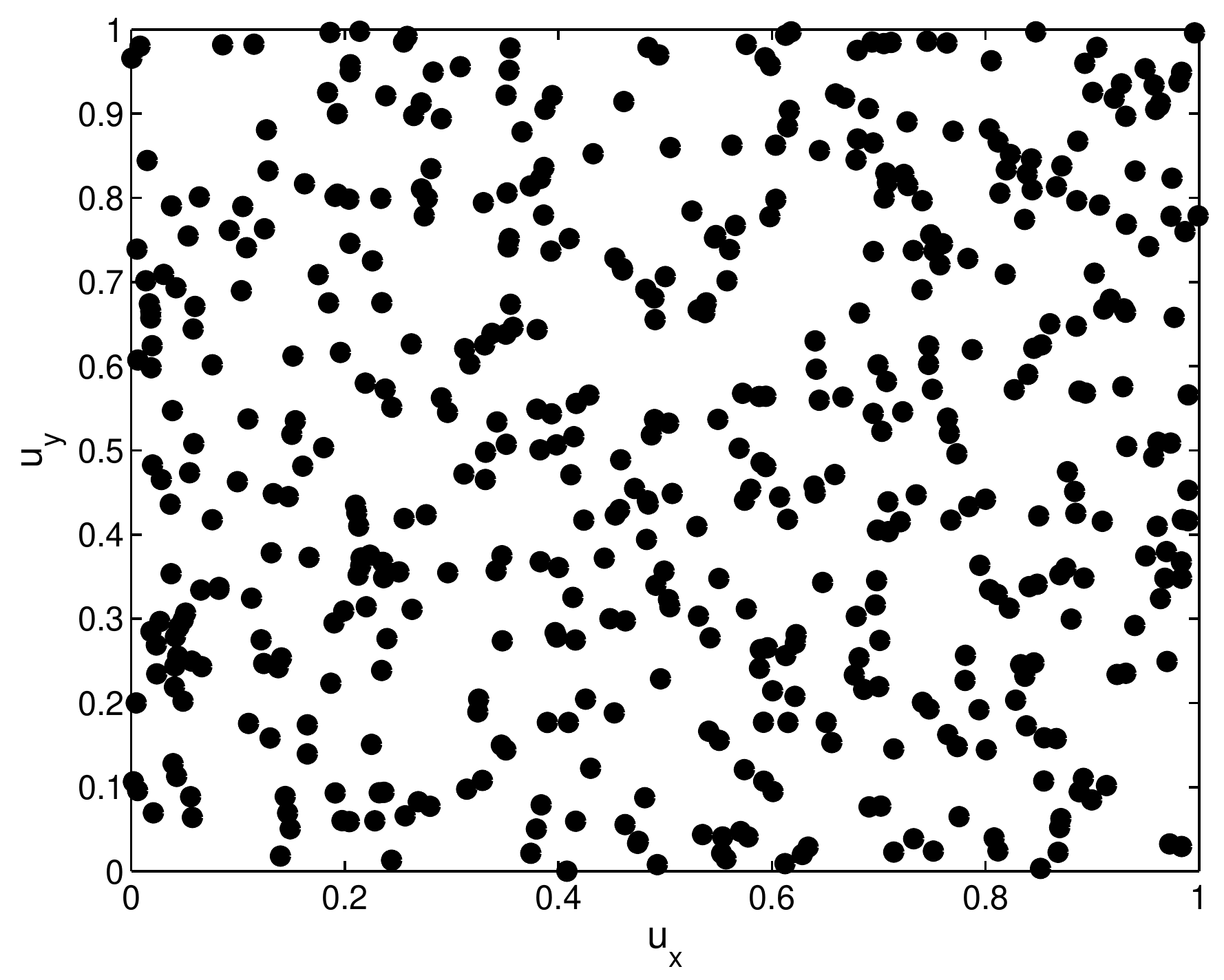}}
\subfigure[]{\includegraphics[width=.24\textwidth]{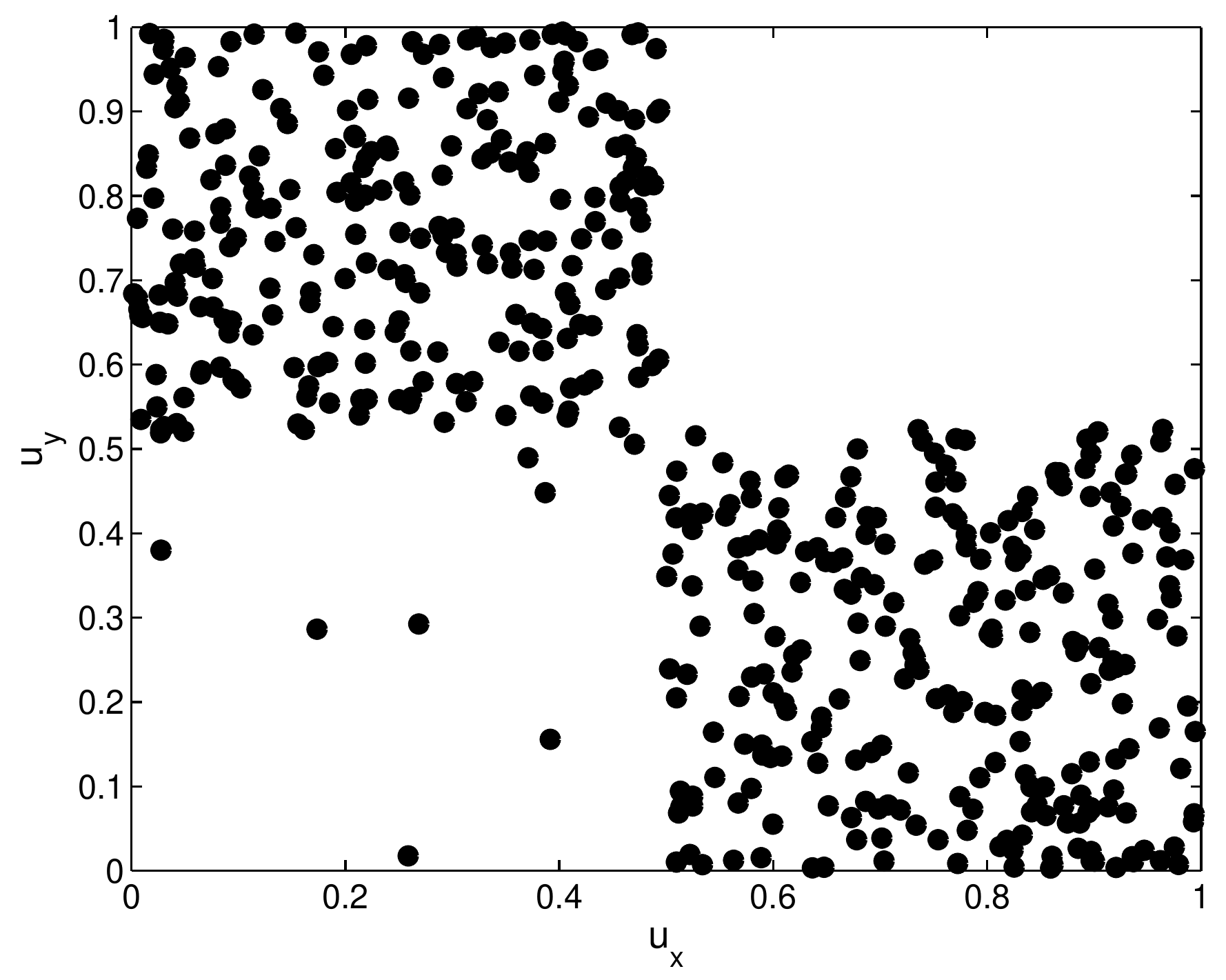}}
\subfigure[]{\includegraphics[width=.24\textwidth]{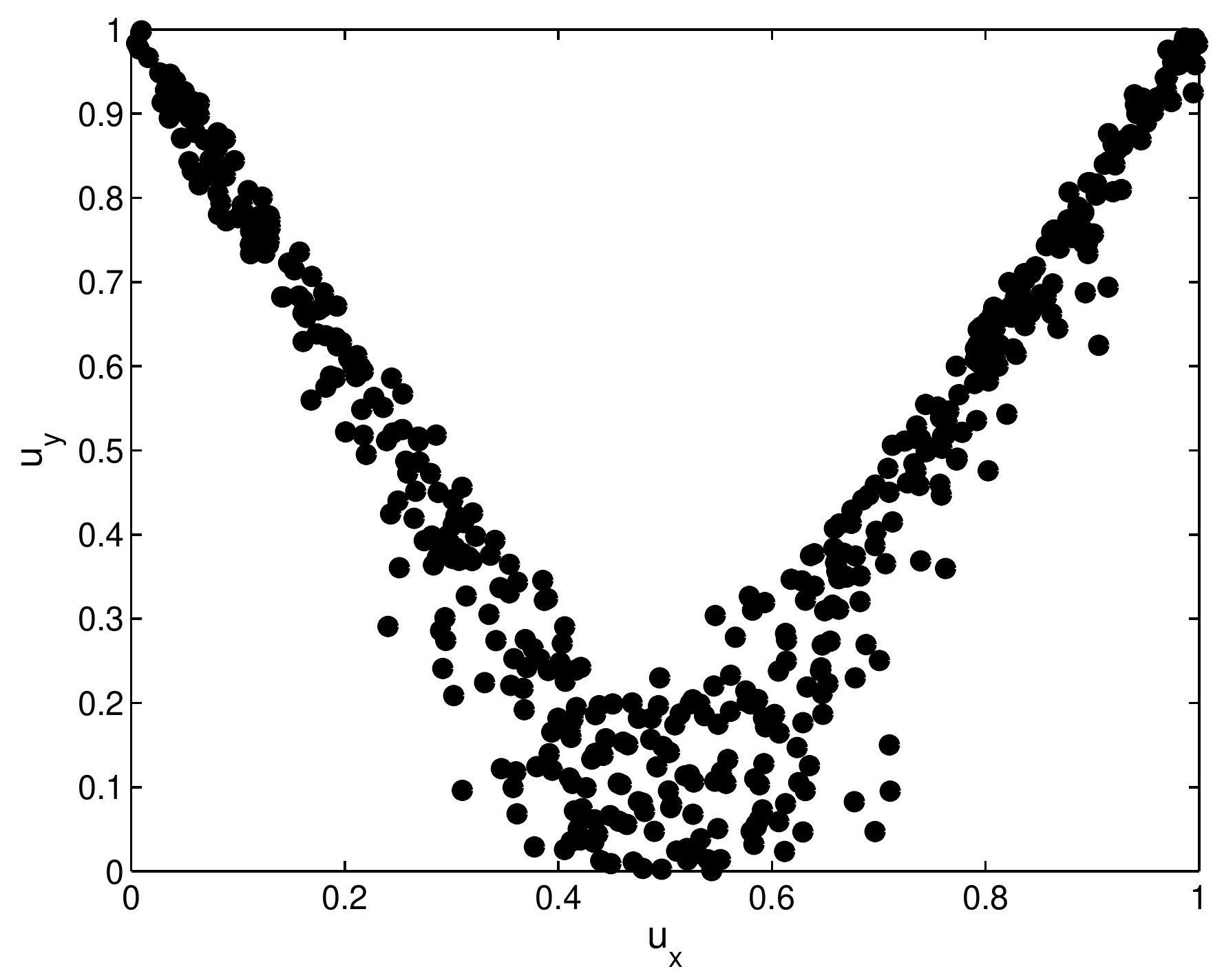}}
\subfigure[]{\includegraphics[width=.24\textwidth]{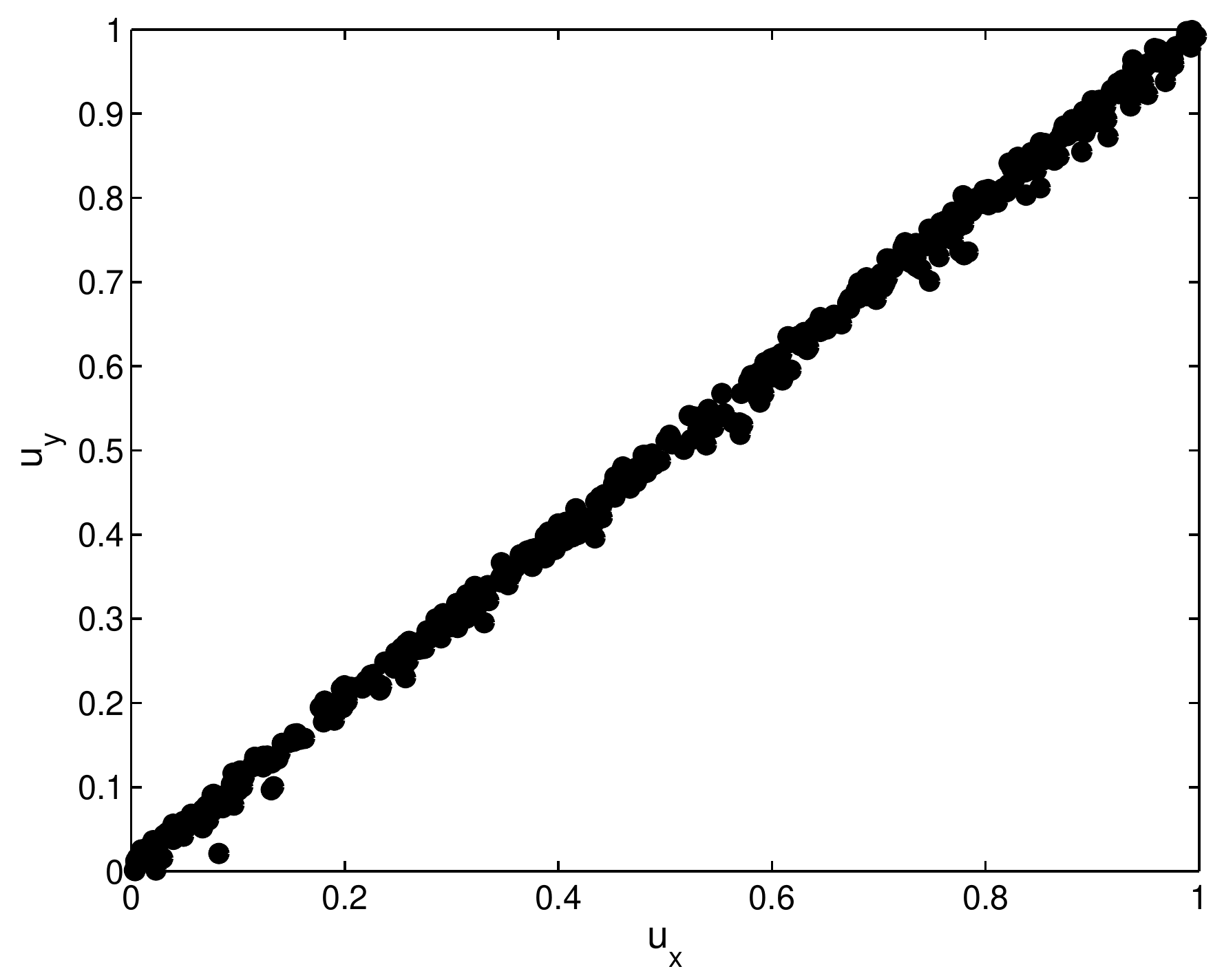}}
\caption{Examples of the Plackett-Luce copula for different functions $\lambda_\beta$. The top figures (a-d) plot the different functions $\lambda_\beta$. The bottom figures (e-h) represent samples from the copula $C_{\lambda_\beta}(u_x,x_y)$, where $X\in[0,1]$ and $F_X$ is uniform.}
\label{fig:copulas}
\end{figure}

The Plackett-Luce copula takes the following form
\begin{align}
C_{\lambda_{\beta}}(u_1,u_2)=u_1-\int_{\omega=0}^{u_1} \exp\left (-\lambda_\beta(\omega) F_Z^{-1}(u_2) \right )d\omega.
\end{align}
Figure~\ref{fig:copulas} shows illustration of the copula for different functions $\lambda_\beta$.

The conditional distribution function can then be expressed as
\begin{align*}
F_{Y\mid X=x}(y) &=  1-\exp(-\lambda_\beta(x) F^{-1}_Z(F_Y(y))).
\end{align*}

Given $\lambda_\beta$, the random variables $Y|X=x$ are stochastically ordered. For $x_1, x_2$ such that $\lambda_\beta(x_1) \leq \lambda_\beta(x_2)$

\begin{align*}
F_{Y\mid X=x_1}(y) \leq  F_{Y\mid X=x_2}(y) \;\; \forall y \in \mathbb{R}.
\end{align*}

If $F_Y$ has a density with respect to Lebesgue measure, $f_Y$, then we can use a change of variables to calculate the conditional density  as follows:

\begin{align*}
f_{Y\mid X=x}(y) &= f_Y(y) \frac{f_{Z\mid X=x}(z(y))}{f_Z(z(y))}\\
&= f_Y(y) \frac{f_{Z\mid X=x}(F_Z^{-1}(F_Y(y)))}{f_Z(F_Z^{-1}(F_Y(y)))}\\
&= f_Y(y) \frac{\lambda_{\beta}(x)\exp[-\lambda_{\beta}(x) F_Z^{-1}(F_Y(y))]}{\int_{\mathcal X} \lambda_{\beta}(x')\exp[-\lambda_{\beta}(x') F_Z^{-1}(F_Y(y))] dF_X(x')}.
\end{align*}

It can be seen from this representation that the conditional density of $Y_i$, given $X_i$ is simply the marginal density of $Y_i$, re-weighted across its quantiles ($F_Y(y)$) by a function of $X_i$.

\bigskip
We end the construction of the model by assuming a prior over the finite-dimensional parameter $\beta$ and Bayesian nonparametric prior over the marginal distributions $F_X$ and $F_Y$
\begin{align}
\beta&\sim\pi_\beta\\
F_Y &\sim \mathbb P_Y\\
F_X &\sim \mathbb P_X
\end{align}
where $\pi_\beta$ is some parametric prior and $\mathbb P_X$ and $\mathbb P_Y$  may be a Dirichlet process mixture or a P\'olya tree prior, as described in Section~\ref{sec:densityestimation}.

\section{Approximations for posterior inference and prediction}
\label{sec:inference}

Assume that both $F_X$ and $F_Y$ admit a density with respect to Lebesgue measure, noted $f_X$ and $f_Y$. The unknown quantities for our regression model are therefore $(f_Y, \beta,f_X)$. Given data $(x_{1:n},y_{1:n})$, where $x_{1:n}=(x_1,\ldots,x_n)$ and $y_{1:n}=(y_1,\ldots,y_n)$,  we have the following likelihood:
\begin{align}
L(f_Y, \beta,f_X;(x_{1:n},y_{1:n})) = \prod_{i=1}^n f_Y(y_i) \frac{\lambda_{\beta}(x_i)\exp[-\lambda_{\beta}(x_i) F_Z^{-1}(F_Y(y_i))]}{\int_{\mathcal X} \lambda_{\beta}(x')\exp[-\lambda_{\beta} (x') F_Z^{-1}(F_Y(y_i))] dF_X(x')} f_X (x_i).
\end{align}
Inference could proceed using numerical methods such as MCMC but for large datasets this is cumbersome. Hence we consider here a Bayesian composite marginal likelihood approach~\citep{Lindsay1988,Cox2004,Varin2011,Pauli2011,Ribatet2012} that we show offers computational tractability and the use of standard Bayesian methods. Define $y^*_{1:n}$ to be $y_{1:n}$ ordered from lowest to highest, and let $\nu_{1:n}=(\nu_1,\ldots,\nu_n)$ be a vector representing the order of $y_{1:n}$, so that $y^*_i = y_{\nu_i}$. Then we can re-write our data $\{y_{1:n}, x_{1:n}\}$ equivalently as $\{y^*_{1:n}, \nu_{1:n}, x_{1:n}\}$. Now let $L_C$ denote the composite marginal likelihood based on $\{y^*_{1:n}\}$ and $\{\nu_{1:n}, x_{1:n}\}$. That is the product of the likelihood terms associated with each of these terms:

\begin{align}
L_C(f_Y, \beta,f_X; \{y_{1:n}, x_{1:n}\}) &= L(f_Y, \beta,f_X; \{y^*_{1:n}\})\times L(f_Y, \beta,f_X; \{\nu_{1:n}, x_{1:n}\})\nonumber \\
&= n! \left [\prod_{i=1}^n f_Y(y_i)\right ] \times \left [\prod_{i=1}^n \frac{\lambda_\beta(x_{\nu_i})}{\sum_{j\geq i} \lambda_\beta(x_{\nu_j})}\right ] \times \left [\prod_{i=1}^n f_X(x_i)\right ].
\label{eq:compositelikelihood}
\end{align}

We can see that this composite likelihood approach factors the likelihood into separate terms involving $f_Y, \beta$ and $f_X$, leading to the following pseudo posterior distribution

%
%
%

\begin{align}
\pi_C(f_Y, \beta, f_X \mid \{y_{1:n}, x_{1:n}\}) = \pi_C(f_Y|y_{1:n}^\star) \pi_C(f_X|x_{1:n}) \pi_C(\beta|\nu_{1:n},x_{1:n})
\label{eq:pseudopost}
\end{align}
Inference over the parameters $f_Y, \beta, f_X$ can thus be carried out independently under the composite likelihood approach. Standard software for Bayesian nonparametric univariate density estimation can be used for $f_Y$ and $f_X$, and software for fitting Plackett-Luce/Cox proportional hazard can be used for fitting $\beta$. Overall the advantages of the approximate composite likelihood approach include computational tractability and scalable inference using standard software, hence good numerical reproducibility, and high interpretability as the components in the composite likelihood have explicit form and meaning. This latter point aids in prior elicitation as it allows the analyst to separate out and represent their beliefs on the marginal distributions, which are simpler to specify than the full conditionals, and then consider the dependence given the marginals.

The Bayesian composite likelihood approach has attracted some attention over recent years~\citep{Pauli2011,Varin2011,Ribatet2012}. In particular, \cite{Ribatet2012} considered two adjustements to the marginal likelihood approach in order to retain some of the desirable properties of the usual likelihood. However, their adjustments apply to a specific form of composite likelihood, where it factorizes as a product of composite likelihoods for each observation: $L^{total}_c(y|\theta) = \prod_{i=1}^n L_c(y_i|\theta)$ where $ L_c(y_i|\theta)$ is the composite likelihood for observation $i$. Our composite likelihood approach does not fit in this framework, as we do not have this product form over the observations, and we cannot therefore apply the adjustments suggested by \cite{Ribatet2012}. Extending the adjustment of \cite{Ribatet2012} to our framework is an interesting direction, but beyond the scope of this article.

\subsection{Asymptotics for the marginal composite posteriors}

Consider first the pseudo-posterior for $f_Y$:

\begin{align*}
\pi_C(f_Y \mid \{y_{1:n}, x_{1:n}\}) &\propto \pi(f_Y) L_C(f_Y; \{y_{1:n}, x_{1:n}\})  \\
&\propto  \pi(f_Y) \prod_{i=1}^n f_Y(y_i).
\end{align*}

So our pseudo-posterior is exactly the posterior based on the i.i.d sample $\{y_{1:n}\}$, where $y_{1:n} \sim F_Y$. This is the standard setting for posterior inference, so we can apply consistency results from Bayesian nonparametric inference for $F_Y$, see for example \cite{ghosal2013fundamentals}. The same is true for $f_X$. Now consider the log-linear form for $\lambda$: $\lambda(x) = \exp(-\beta x)$. Then, we have the pseudo-posterior:

\begin{align*}
\pi_C(\beta \mid \{y_{1:n}, x_{1:n}\}) &\propto \pi(\beta) L_C(\beta; \{y_{1:n}, x_{1:n}\})  \\
&\propto  \pi(\beta) \prod_{i=1}^n \frac{e^{\beta x_{\nu_i}}}{\sum_{j\geq i} e^{\beta x_{\nu_j}}}.
\end{align*}
This is exactly the posterior considered by \citet{kim2006} in a different setting where a Bernstein-Von Mises theorem is proven, which can be applied here.

\subsection{Posterior predictive}

%
%
%
%
%
We can use simulation methods such as MCMC to easily generate samples $\{F_Y^{(j)}, \beta^{(j)}\}_{j=1}^m$ from the pseudo-posterior~\eqref{eq:pseudopost}; the predictive distribution can then be approximated by
$$p(y' \mid x',\{y_{1:n},x_{1:n}\}) \simeq \frac{1}{m}\sum_{j=1}^m  p(y' \mid x', \beta^{(j)}, F_Y^{(j)}). $$

To simulate from this distribution, we can use the forward generating process of our model, given $X=x'$:

\begin{align}
Z'|X'=x' &\sim \text{Exp }(\lambda_\beta(x'))\\
Y' &= F_Y^{-1}( F_Z(Z'))\label{eq:YgivenZ}.
\end{align}

In many applications, modeling $F_X$ might be cumbersome, and not the primary object of interest. In this case we propose to use an empirical Bayes approach by setting $F_X = \hat{F}_X$ at the empirical CDF. So, to generate a posterior predictive sample, given a posterior sample $\{F_Y^{(j)}, \beta^{(j)}\}_{j=1}^m$, Eq.~\eqref{eq:YgivenZ} becomes:

$$Y'^{(j)} = F_Y^{-1 (j)}\left (1- \frac{1}{n} \sum_{i=1}^n  e^{-Z'^{(j)} \lambda_{\beta^{(j)}} (x_i)}\right )$$
where we note that $Z'^{(j)}$ is conditional on $X'=x'$, and the CDF inversion is tractable, depending on the form of $F_Y$. Alternately one can use Monte Carlo to draw samples from the predictive, which is trivial when $F_Y$ can be sampled from. Some particular examples are discussed in Appendix A.

\section{Illustrations}
\label{sec:appli}

In this Section we apply our method to two examples. The first is a simulation example where we generate from a multi-modal conditional and explore the ability of our method to fit the data. The second is a large real-world application in the regression analysis of US Census data.


\subsection{Simulation example}

\begin{figure}[h]
\centering
\subfigure[Data]{\includegraphics[width=0.49\textwidth]{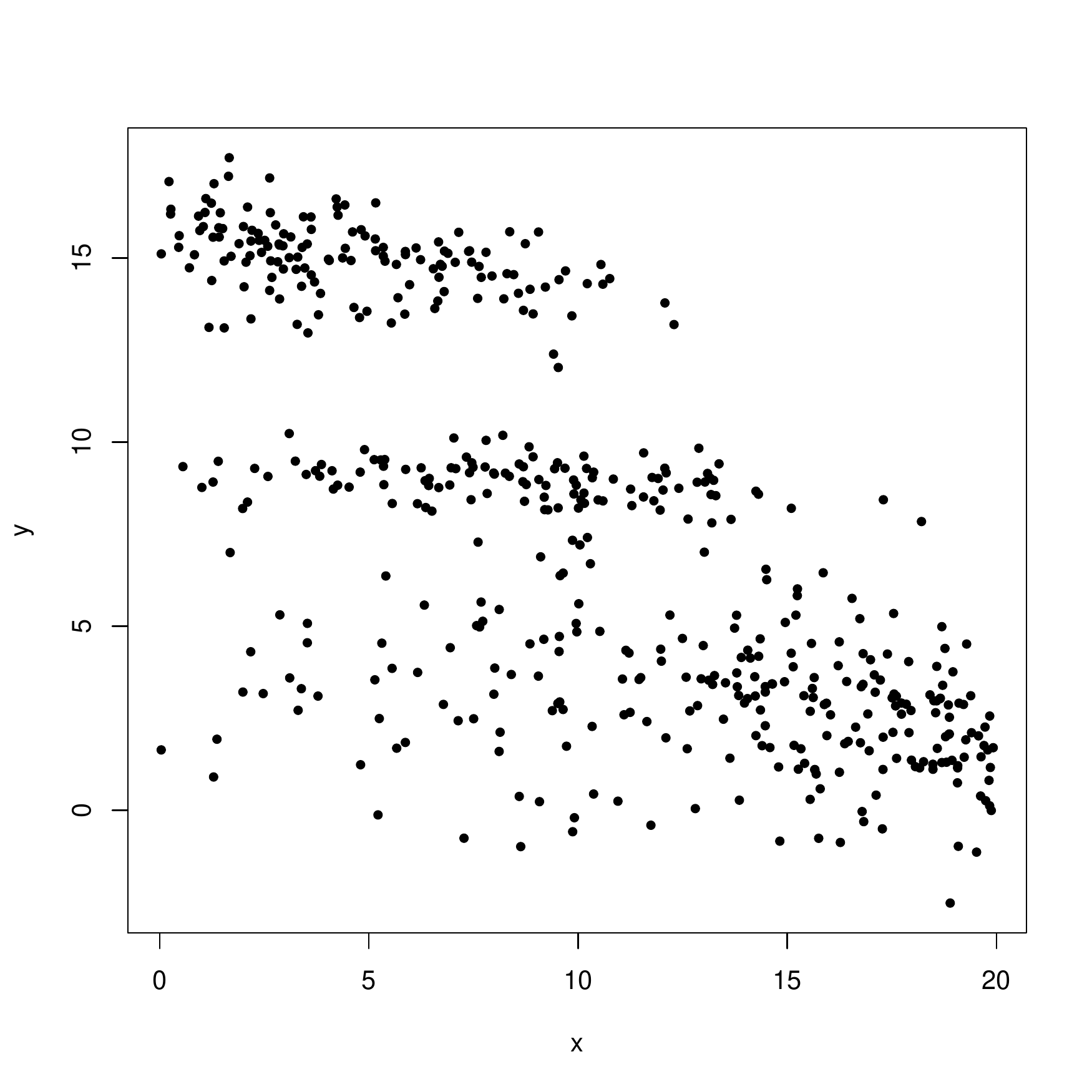}}
\subfigure[80 \% highest posterior predictive intervals]{\includegraphics[width=0.49\textwidth]{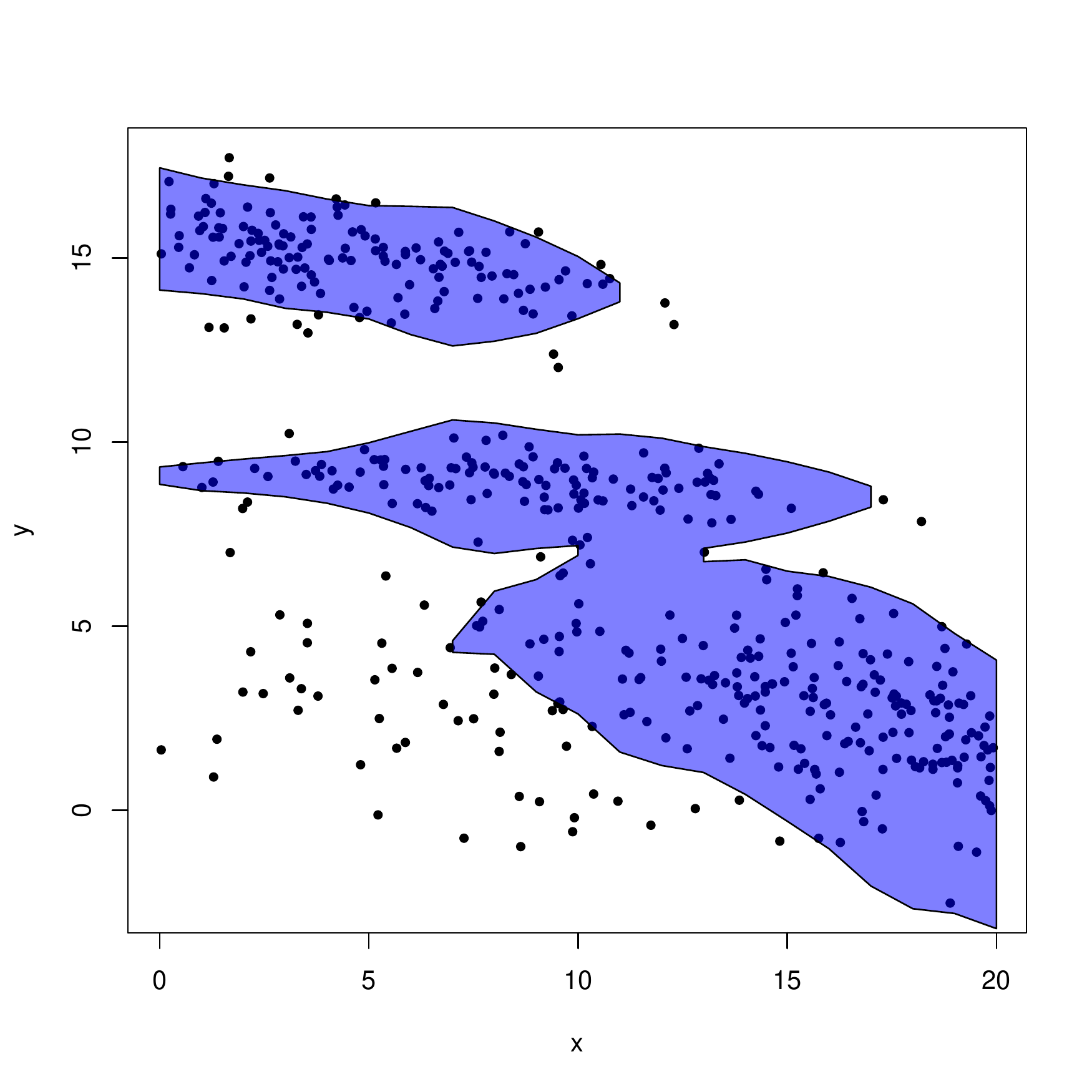}}
\caption{(a) Data simulated from the model with mixture of three Gaussians marginal distribution for $Y$. (b) 80\% highest posterior density intervals of the predictive distribution at each value of $x$.}
\label{fig:sim_data}
\end{figure}

In this section we apply the model to a dataset simulated from our model to consider how well we can recover known dependence. The marginal distribution of $Y$ is set to a mixture of three Gaussian distributions, with means 3, 9 and 15, standard deviations of 2, 0.5 and 1 with mixture weights of 0.5, 0.2 and 0.3 respectively. $\beta$ is set to 0.25, with $\lambda_{\beta}(x)=\exp(\beta x)$. $X\sim\Unif(0,20)$ and $n=500$. The data is shown in Figure \ref{fig:sim_data}(a).

Clearly any type of linear or non-linear regression with a parametric noise distribution will be inappropriate here. The conditional distribution of $Y$ given $x$ is multi-modal, rendering many popular regression models inappropriate.

We compared our approach to a linear dependent Dirichlet process mixture of normals (LDDPM) \citep{DeIorio2004}, using the R package DPpackage~\citep{dppackage, jara07}.  This model specifies that

\begin{align*}
&Y_i \mid x_i \sim  \int \mathcal N(y_i; x_i \beta, \sigma^2) G(d\beta, d\sigma^2)\\
&G \mid \alpha, \mu_b, s_b \sim  \DP(\alpha G_0)
\end{align*}

where $G_0= \mathcal N(\mu_b, s_b) \Gam( \tau_1/2, \tau_2/2)$ and

\begin{equation*}
s_b \mid \nu, \psi \sim IW(\nu, \psi)
\end{equation*}
with  $\alpha=1, \mu_b=(9,0)^T, \nu=4, \tau_1=1, \tau_{2}=2, \psi = \bigl(\begin{smallmatrix}
1&0\\ 0&1 \end{smallmatrix} \bigr)$ and
 $s_b= \bigl(\begin{smallmatrix} 36&0\\ 0&36 \end{smallmatrix} \bigr) $.

We apply our model, modeling the marginal as a Dirichlet Process mixture of Gaussian distributions using $\alpha=1$ and a normal-inverted-Wishart distribution for the base measure. That is, our base measure $G_0(\mu, \sigma^2) = \mathcal N(\mu \mid \mu_1,  \frac{\sigma^2}{\kappa_1}) IW (\sigma^2 \mid \nu_1, \psi_1)$, where $\mu_1 = 9, \kappa_1=0.5, \nu_1=4$ and $\psi_1=1$. A Gaussian prior centered at $0$ with unit variance is used for $\beta$.

\begin{figure}[h]
\centering
\subfigure[]{\includegraphics[width=0.49\textwidth]{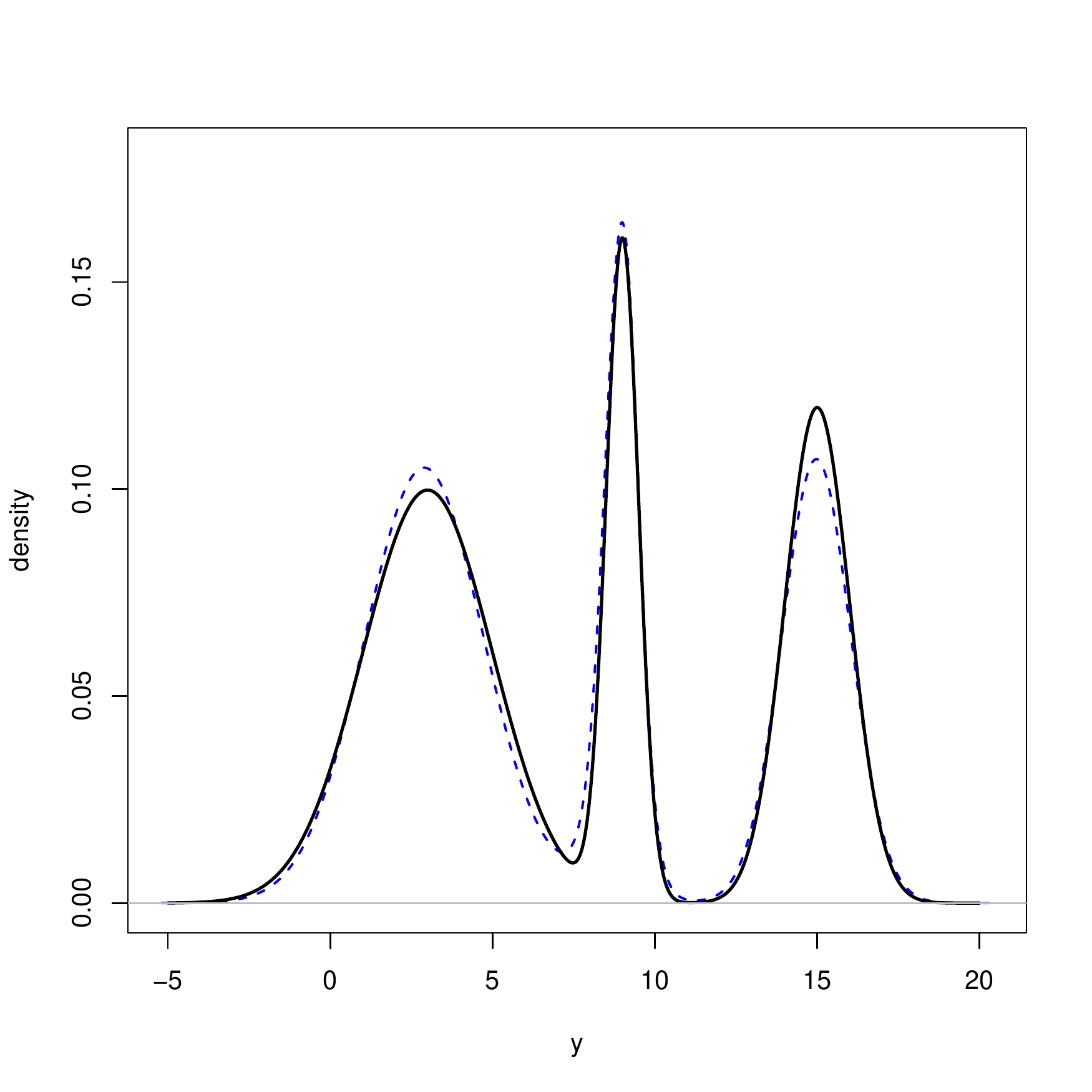}}
\subfigure[]{\includegraphics[width=0.49\textwidth]{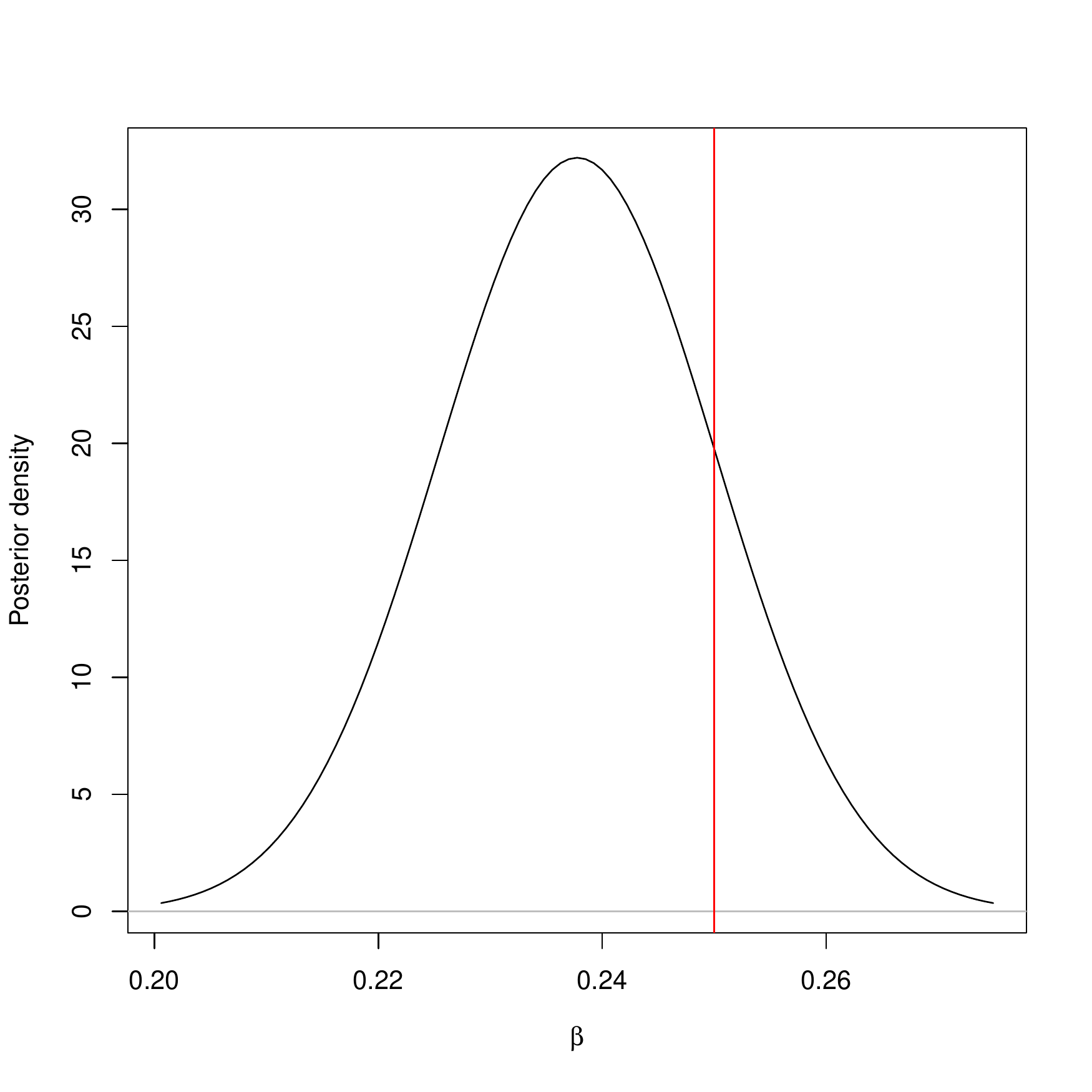}}
		
\caption{(a) The posterior predictive marginal for $y$ under our model in blue, compared to the actual sampling distribution in black. (b) The posterior distribution for $\beta$, compared to the true value of 0.25 marked in red.}
\label{fig:marg}
\end{figure}

In Figure \ref{fig:sim_data}(b) the simulated data is shown, with the 80\% highest posterior density (HPD) intervals of the predictive distribution at each value of $x$. Qualitatively we see that the model can capture the nonlinearities in the data and demonstrates the flexibility to model the multi-modal conditional response. In Figure \ref{fig:marg} we show the predictive marginal, $\hat{F}_Y$ and the posterior distribution for $\beta$. Clearly the marginal distribution for $Y$ is very well recovered from the data. This parameterization of the model in terms of the marginal distribution for the response allows this to be estimated from the complete dataset, without reliance on other aspects of the model. The strength of information available is apparent in the quality of the fit to the sampling distribution. The posterior for the parameter $\beta$ shows reasonable support around the true value, being slightly pulled towards 0 by the prior.

\begin{figure}[h]
\centering
\subfigure[$x=5$]{\includegraphics[width=0.49\textwidth]{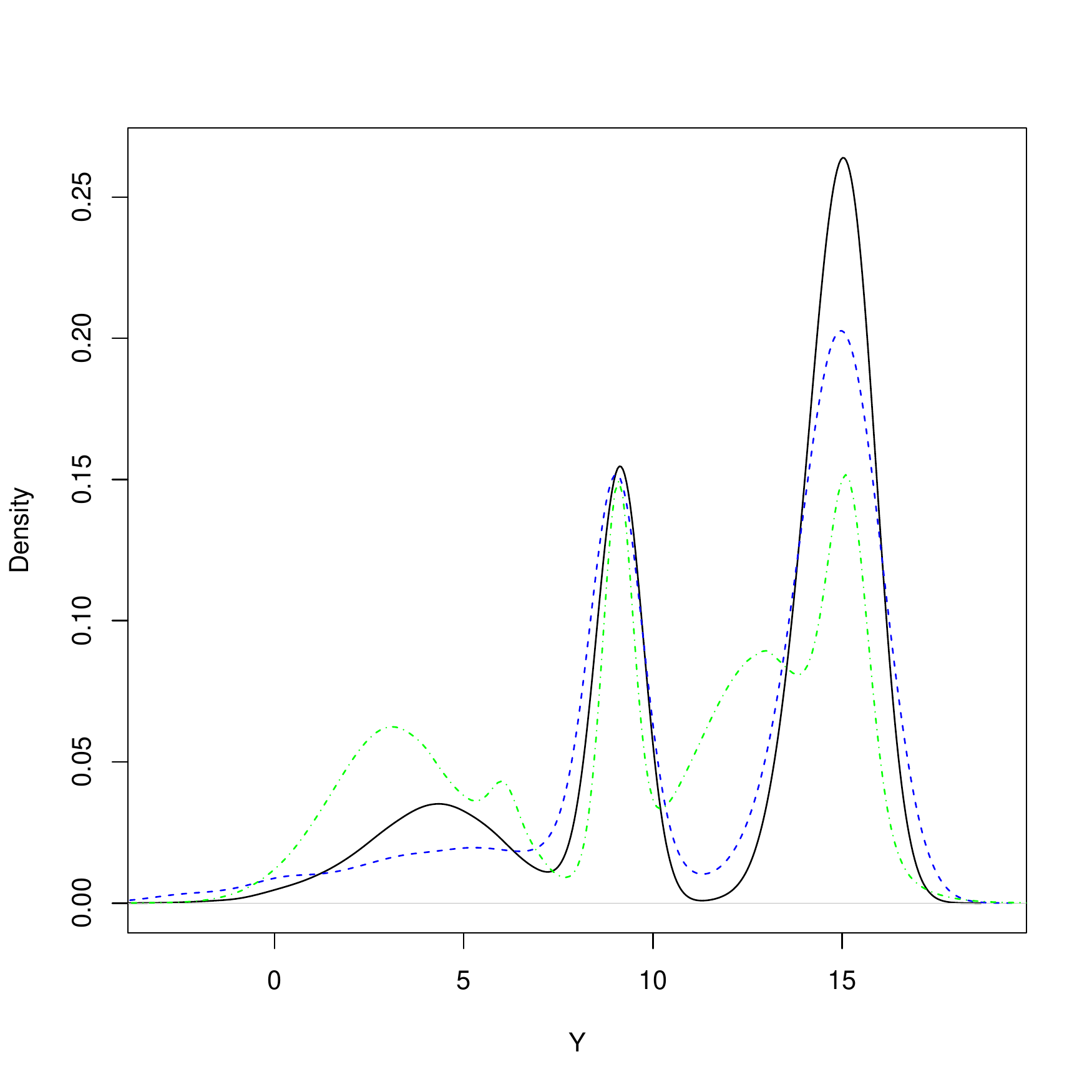}}
\subfigure[$x=12$]{\includegraphics[width=0.49\textwidth]{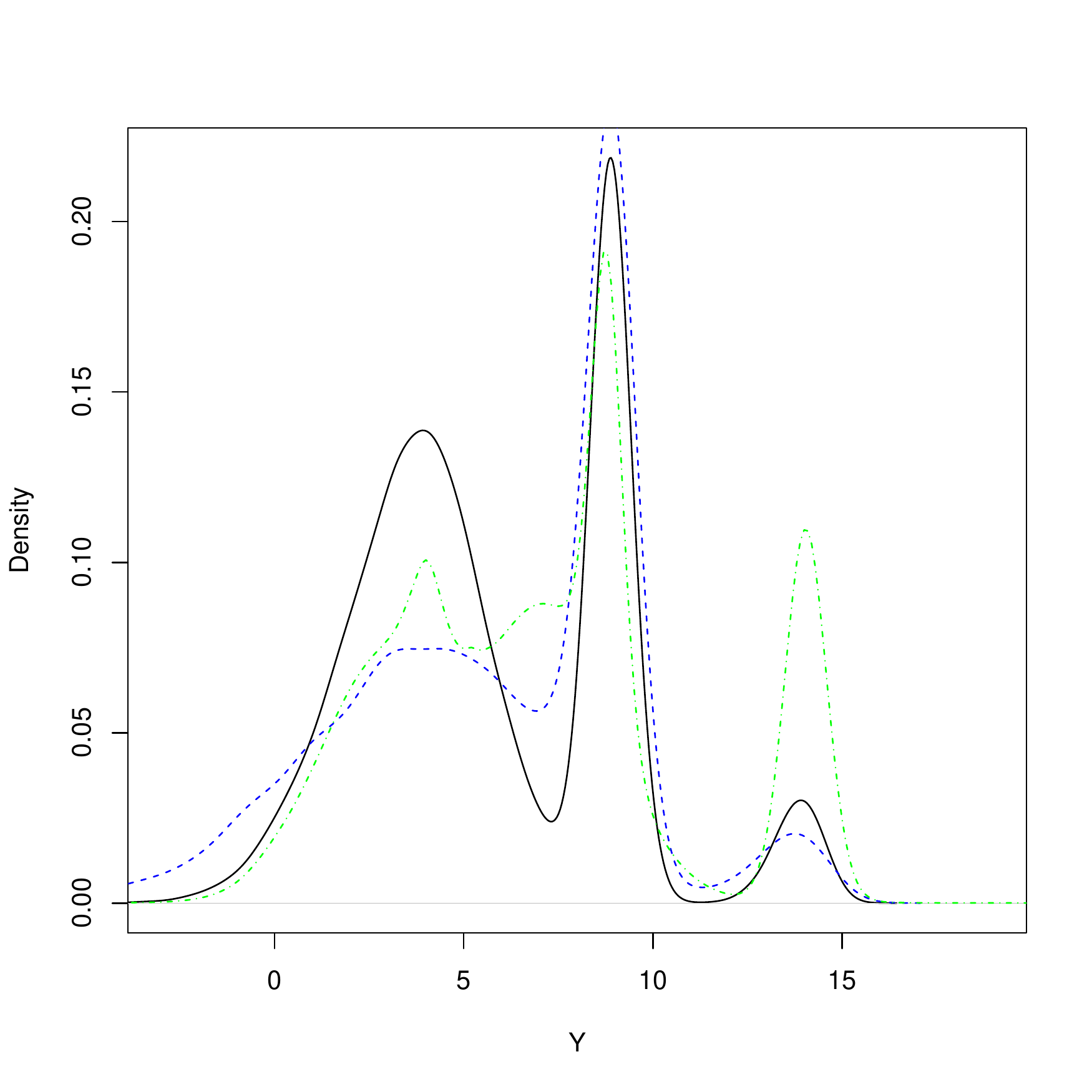}}		
\caption{Predictive densities for (a) $x=5$ and (b) $x=12$. The true predictive is shown in black, the predictive distributions under our model in blue, and the predictive under the linear DDP mixture in green.}
\label{fig:pred}
\end{figure}

We can further inspect how these come together in the posterior predictive conditional distribution for $Y$ given $x$. Consider this distribution for $x=5$ and $x=12$, for both our model and the linear DDP mixture model, as shown in Figure \ref{fig:pred}. Again, our model provides a reasonable fit. The predictive distribution is not as accurate as the marginal distribution for $Y$, but this is to be expected, since the conditional distribution is a product of the whole model, compounding uncertainties from both $\beta$ and the marginal distribution for $y$. Nonetheless, the fit is good and noticeably better than the flexible linear DDP mixture, as you would expect, given that the sampling distribution is within the support of our model. Concretely, the L1-distance between the estimated conditional and the true conditional distribution can be calculated in each case. When $x=5$ the distance to our prediction is 0.00869, whereas the distance to the linear DDP is 0.0214, and when $x=12$ the distance to our prediction is 0.0127 and the distance to the linear DDP is 0.0146.

A point of note is that these posterior predictive plots are smoothed kernel density estimates of MCMC samples. Therefore, Gaussian shapes are slightly exaggerated. Whilst not entirely clear from the plot, both our predictive and the sampling distribution comprise of slightly skewed Gaussian distributions, since the conditional distribution is the marginal distribution for $Y$ weighted across the quantiles.


\begin{figure}[h!]
\centering
\includegraphics[width=0.5\textwidth]{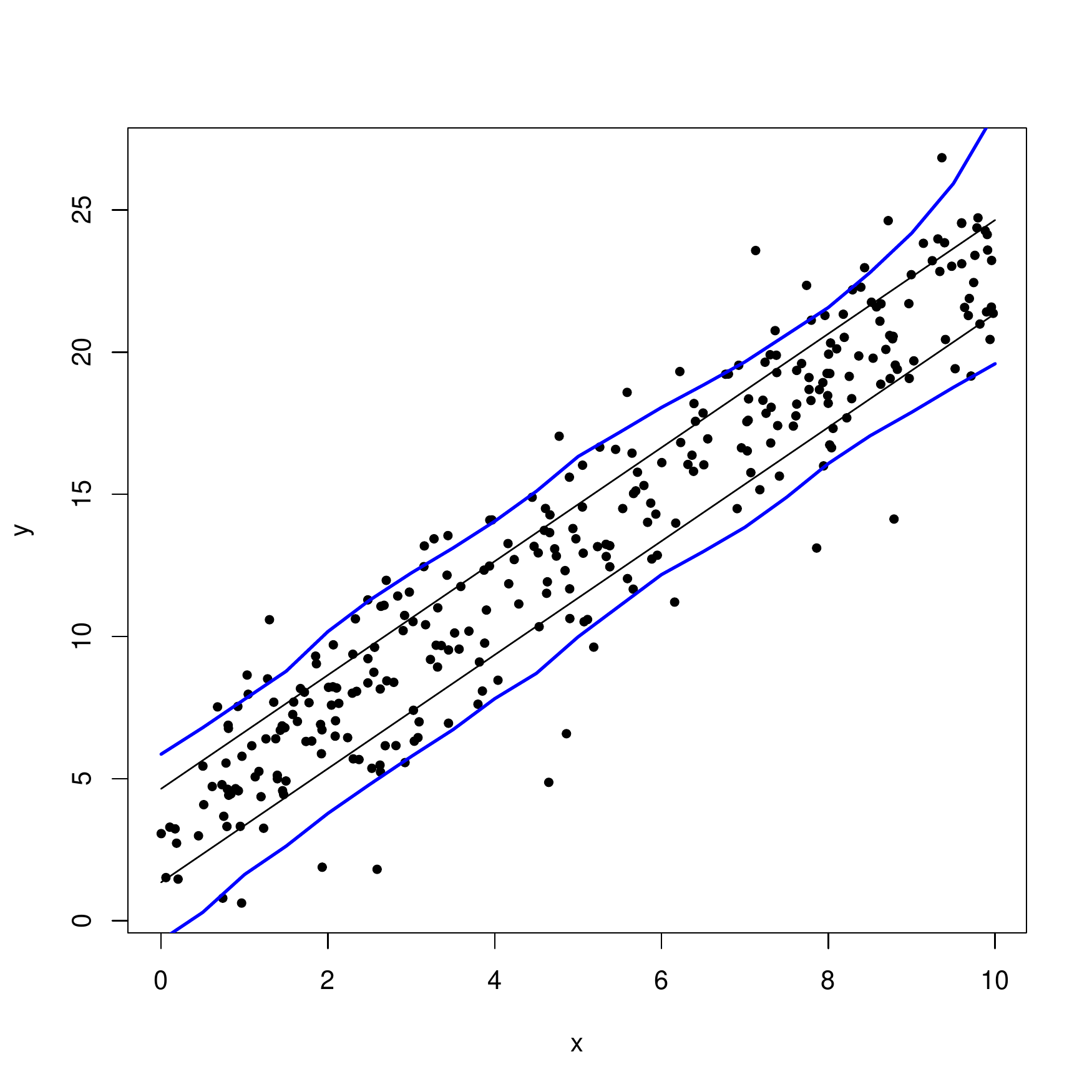}
\caption{Simulated data from a linear model with Gaussian residuals (black dots). 80\% HDP intervals of the predictive distribution of our model at each value of $x$ are represented in blue, and true HDP interval in black }
\label{fig:fitlin}
\end{figure}

To illustrate that the model is capable of modeling a range of distributions, we consider data sampled from a Gaussian linear model. The covariates are simulated uniformly on $[0,10]$, with $Y \sim N(3 + 2 x, 2)$ and $n=300$. We use our model, modeling the marginal for $Y$ with a P\'olya tree prior whose partition is set on a Gaussian distribution with mean 12.5 and standard deviation 6, and $\alpha_{\epsilon_1 \ldots \epsilon_m} = m^2$. A Gaussian prior centered at 0 with variance $8$ is used for for $\beta$. The posterior predictive 80\% HPD intervals display a reasonable fit of the linear data, shown in Figure \ref{fig:fitlin}. The variance seems slightly inflated, but this is a consequence of the large support of the model.

\subsection{US Census application}

We apply the methodology to a regression task using US census data\footnote{\url{http://www.census.gov/acs/www/data_documentation/pums_data/}} for personal annual income.

We use the American Community Survey data from 2013, which comprises of responses to questions on the survey given to a $1\%$ sample of the US population. Since we are interested in income, the subset of $1,371,401$ employed civilians over the age of 16 is used. We have used a relevant, linearly independent subset of the data as covariates, excluding highly informative questions such as occupation, which would almost completely explain the response.  This leaves 15 explanatory variables, 10 of which are categorical variables, some of which have many levels. The result is a $1,371,401 \times 114$ design matrix.

The covariates are: US state (Texas as a baseline), weight, age, class of worker (employee of private for-profit company as a baseline), travel time to work, means of transportation to work (works from home as a baseline), language other than English spoken at home (no as a baseline), marital status (married as a baseline), educational attainment (regular high school diploma as a baseline), gender (male as a baseline), hours worked a week, weeks worked last year, disability status (without a disability as a baseline), quarter of birth (first quarter as a baseline), and world area of birth (United States of America as a baseline).

\begin{figure}[h]
\centering
\includegraphics[height=0.3\textheight]{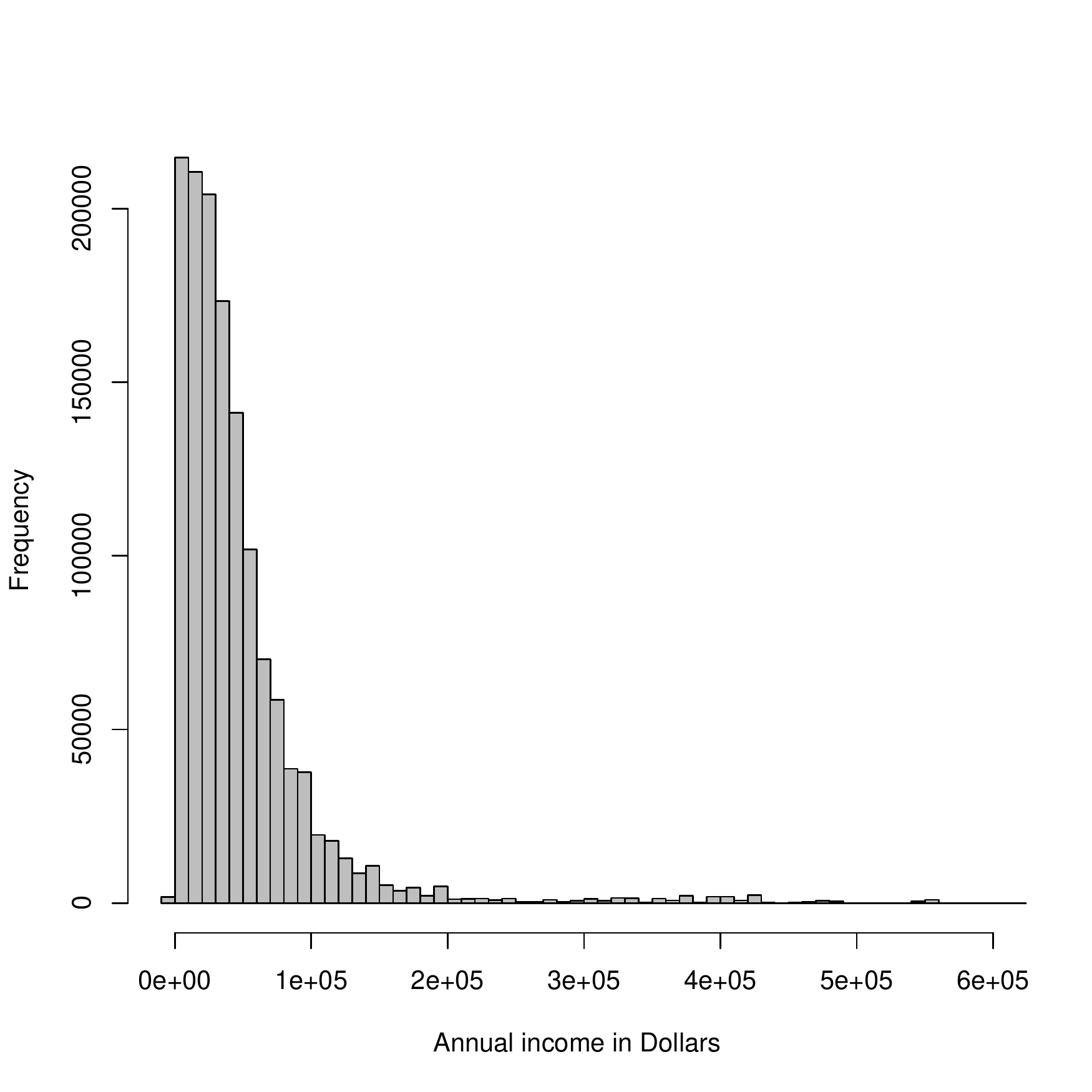}
\includegraphics[height=0.3\textheight]{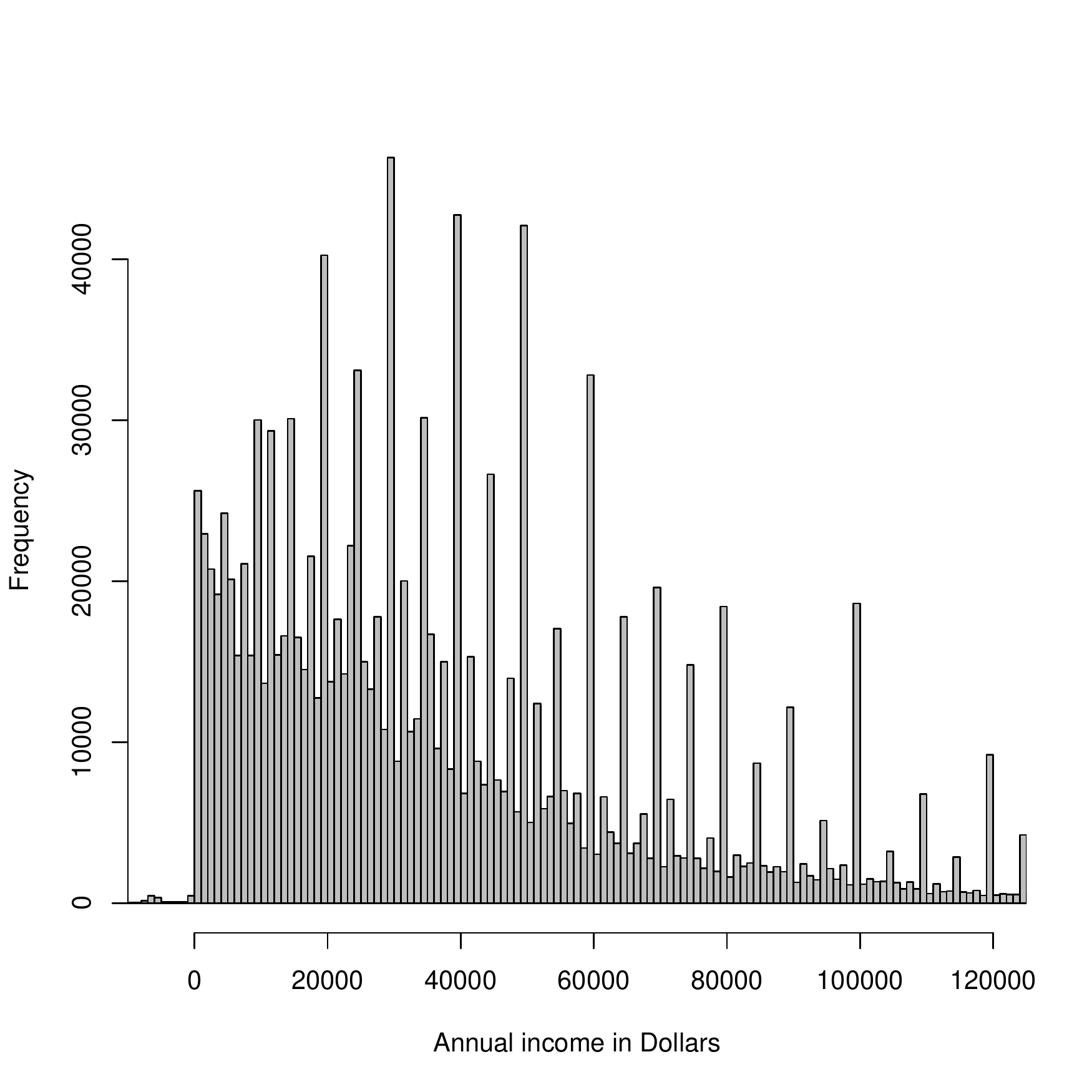}
\caption{Histograms of annual income on different scales. Right hand plot is zoomed in on incomes up to 120,000.}
\label{fig:census}
\end{figure}

The levels of annual income shown in Figure (\ref{fig:census}) can be seen to be heavy tailed, which requires a flexible model to capture. Another noticeable feature of the data is that the income levels are discontinuous, with large spikes in frequency at particular income levels. This could in part be due to standardized salary structures resulting in certain salary levels becoming common. This motivates the use of a nonparametric approach as it is difficult to imagine how a parametric density could conditionally capture the features shown in Figure (\ref{fig:census}). However, standard Bayesian nonparametric models simply cannot be applied to a problem of this scale. Attempting to apply existing methods in this literature, such as the linear dependent Dirichlet process mixture, failed to run due to the dimensionality and scale of the data.

For the analysis, we consider both the empirical distribution function and a P\'olya tree prior for the marginal distribution of $y$. The  partition of the P\'olya tree is set on the quantiles of  a Gaussian distribution with mean 35,000 and standard deviation 20,000, and $\alpha_{\epsilon_1 \ldots \epsilon_m} = m^2$. We use a log-linear regression function $\lambda(x) = \exp(\beta x)$ and place independent Gaussian priors with mean 0 and unit variance on the coefficients in $\beta$.

\subsubsection{Predictive performance}

We compare the out-of-sample predictive performance of our model with three competing non-Bayesian approaches namely, a standard linear regression model, a median regression model and a LASSO\footnote{These models were fitted in R using the functions lm, rq (from the quantreg package) and lars (from the lars package). For LASSO the regularization parameter was chosen using cv.lars. Default settings were used for each.}. For our model we investigated three distinct priors for the marginal distribution of the response: a P\'olya tree centred on a Gaussian, a P\'olya tree centred on a Laplace, and an empirical Bayes approach using the empirical CDF. To compare methods we use repeated random subsets of $1000$ test samples and train each model on the remaining data, with $10$ repeats. Predictive accuracy is judged by mean squared-error (MSE), mean absolute error (MAE) and qualitatively via a qq-plot. To create the qq-plots  we compute the predictive distribution function $F(y | x)$ evaluated at the observed value for each of these test samples. Under the assumption that we have a perfect predictive distribution, these values should be independent uniform random variables. A deviation from this distribution implies a mis-match of the posterior predictive and the actual distribution. We are unable to apply this approach to the median regression model, as it does not provide a predictive distribution and would require fitting the model for a large number of quantiles.  In the case of the linear model we used maximum likelihood estimates for prediction, rather than a fully Bayesian approach. With such a large dataset the strength of any reasonable default prior would be significantly diminished, so this should mimic a Bayesian approach well.

\begin{table}[h]
\centering
\caption{Mean out-of-sample prediction errors with standard deviation of this error after $10$ repetitions}
\begin{tabular}{l |c c}
                   & Mean square error ($10^9$) & Mean absolute error($10^4$)\\
\hline
Empirical model*    & $2.79 \pm 0.51$     & $2.44 \pm 0.15$\\
P\'olya tree (Gaussian)* & $2.81 \pm 0.64$     & $2.41 \pm 0.16$\\
P\'olya tree (Laplace)* & $2.71 \pm 0.52$     & $2.44 \pm 0.14$\\
Linear model       & $2.66 \pm 0.59$    & $2.67 \pm 0.16$\\
LASSO        & $2.99 \pm 0.65$  &  $2.81 \pm 0.16$\\
Median regression       & $2.99 \pm 0.68$    & $2.48 \pm 0.17$\\

\end{tabular}
\label{tab:mse}
\end{table}

\begin{figure}[h]
\centering
\subfigure[Proposed model with P\'olya tree prior]{\includegraphics[height=0.30\textheight]{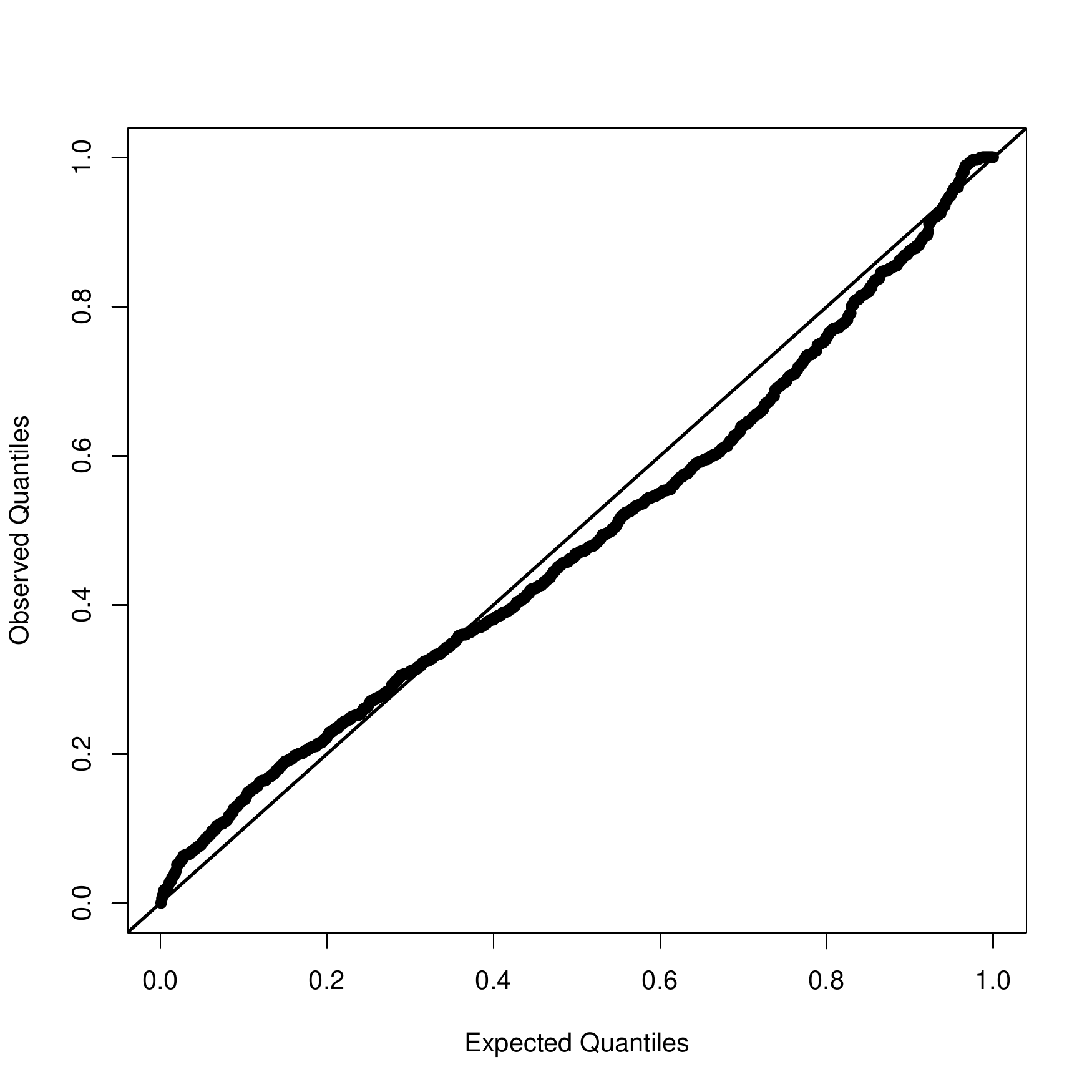}}
\subfigure[Linear model]{\includegraphics[height=0.30\textheight]{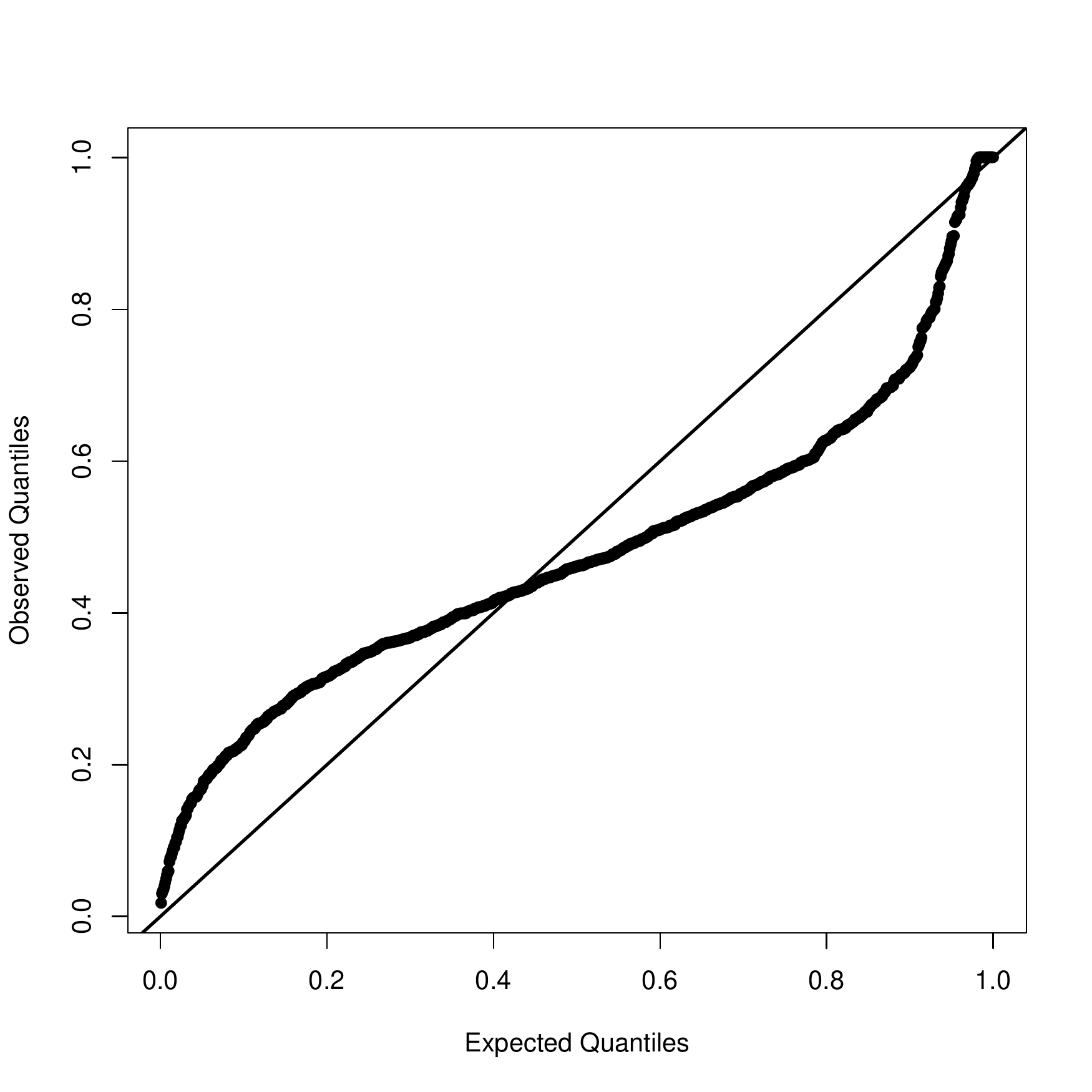}}
\caption{qq-plots (a) under our model using a P\'olya tree prior centred on Laplace for the marginal distribution of the response and (b) using a standard linear model.}
\label{fig:qqplot}
\end{figure}

Summary statistics of predictive fit are shown in Table~\ref{tab:mse}. Perhaps unsurprisingly on such a large data set the linear model targeting the conditional mean does best on MSE but this is at the expense of the median under MAE. In addition, studying the predictive qq-plot in Figure(\ref{fig:qqplot}b) shows the inadequacy of the linear model to provide calibrated predictions. The LASSO performs relatively poorly suggesting most covariates are influential for prediction, whereas the median regression whilst, as expected, provides relative accuracy on the MAE does so at the expense of MSE and as mentioned above suffers from the lack of a fully predictive model. The Bayesian nonparametric methods perform relatively well on both summary measures, with perhaps that based on the Laplace marginal showing greatest accuracy. In Figure(\ref{fig:qqplot}a) we show the predictive qq-plot from this model, demonstrating that the full predictive distribution is captured well. 

 These diagnostics suggest that even non-linear regression models with parametric noise would not provide a satisfactory fit for the data, since the unusual conditional distribution of the response cannot be captured by such models. This highlights the benefit of our nonparametric approach.

We next consider inference for covariate effects. In order to gain a measure of the relevance of each covariate we quantified the concentration of the posterior probability measure away from the prior ``null'' centring of $\beta_j=0$.  To do this we estimated the Bayesian sign-probability from the posterior marginal for each covariate as,
\begin{equation}
PrSign_j = \max\left[ \int_{\beta < 0} \pi(\beta_j | \cdot) d\beta , \int_{\beta >0} \pi(\beta_j | \cdot) d \beta \right]
\label{Bayespval}
\end{equation}
where $\pi(\beta_j | \cdot)$ is the posterior marginal for $\beta_j$. This measures the relative tail area in the posterior marginal laying to the left or right of $0$. A large value of $PrSign$ suggests there is strong evidence against $\beta_j=0$. In certain respects this is akin to a Bayesian marginal version of a p-value, and is trivially calculated from MCMC output, or from normal approximations to the posterior distribution. Table~\ref{tab:psign} shows the most relevant covariates as ranked by this measure.

\begin{table}[h!]
\caption{Top covariate parameters ranked by~\eqref{Bayespval}: the log posterior probability of the parameter being a different sign to the posterior mean. A negative parameter value has a positive effect on income.}
\begin{tabular}{l |c c}
                   & Log probability of different sign & Posterior mean \\
\hline \\
Hours worked a week                &           $  -1.1 \times 10^5 $  &-0.044\\
Weeks worked last year              &          $  -1.0 \times 10^5$   &-0.045\\
Bachelor's degree                          &     $ -4.2 \times 10^4 $  &-0.80\\
Master's degree                               &  $ -4.16 \times 10^4 $  &-1.0\\
Professional degree beyond a bachelor's degree & $ -2.7 \times 10^4$  & -1.4\\
Age                                            & $ -2.6\times 10^4$  & -0.018\\
Female                                         & $ -1.8 \times 10^4$  &  0.35\\
Doctorate degree                          &   $    -1.5 \times 10^4$ &  -1.2\\
Never Married                                  &  $-1.3 \times 10^4$   & 0.37\\
Associate's degree                             &  $ -6.5 \times 10^3$ &  -0.39\\
Travel time to work                               & $-3.2 \times 10^3$ & -0.0035\\
1 or more years of college credit, no degree  &  $  -2.0 \times 10^3$ & -0.18\\
Self employed (incorporated)                    &  $-2.0 \times 10^3$ & -0.30\\
Grade 11 in school                                        &  $-1.8 \times 10^3 $ & 0.38\\
Walks to work                                          &  $-1.5 \times 10^3 $ & 0.36\\
Disabled                                       &  $ -1.3 \times 10^3$ &  0.19\\
\end{tabular}
\label{tab:psign}
\end{table}

Unsurprisingly, hours worked a week and weeks worked last year show high certainly of a positive effect on income. After these, educational achievement measured via degrees unsurprisingly imply higher earnings compared to the regular high school diploma. Since these are part of the same variable it is simple to compare the effects due to these degrees. Despite Bachelor's degree providing the most certainty of a positive effect, a further Professional degree beyond bachelor's has the highest posterior mean effect. The ranking in Table~\ref{tab:psign} reflects the greater evidence in the data for a non-zero Bachelor effect, due to a much higher number of observations of those with Bachelor's degrees, and hence lower variance in the effect size compared with those with a higher degree. There is also strong evidence for Female workers earning less than their male counterparts, as well as increasing income with age and even travel time to work.

\begin{figure}[H]
\centering
\includegraphics[height=0.3\textheight]{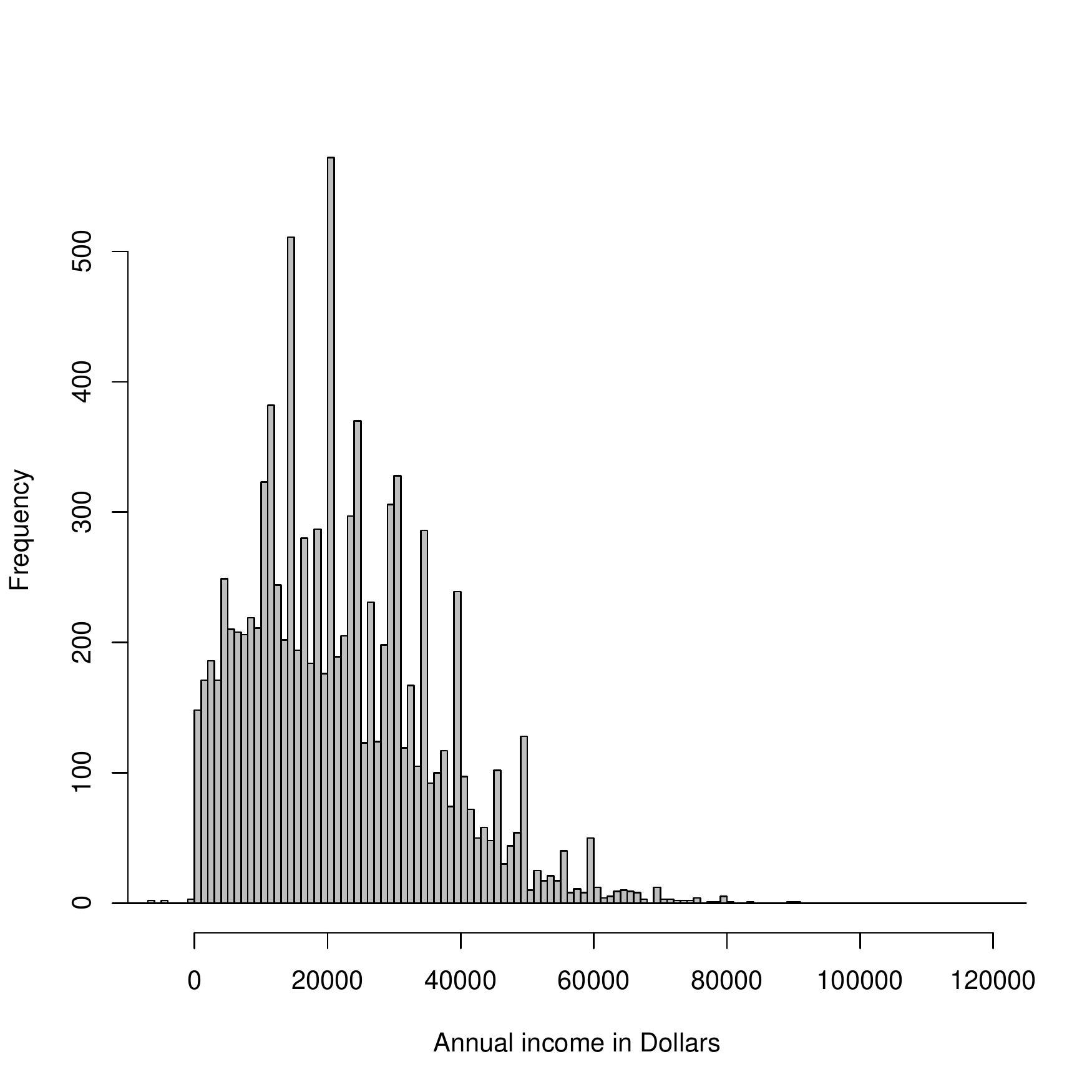}		
\caption{Posterior predictive distribution using a P\'olya tree for the marginal distribution}
\label{fig:lin_qq}
\end{figure}

Finally, we show it is simple to provide the full posterior predictive distribution of annual income of somebody in the test sample, using the P\'olya tree model. We choose as a hypothetical person a 57 year old female from North Carolina, who is self employed, married, 140 lbs bodyweight, who works from home,  speaks English at home, went to college but for less than a year, who usually works 30 hours a week, for 43.5 weeks last year, was born in the first quarter of the year in the USA.
The structure and shape of the posterior predictive, represented in Figure~\ref{fig:lin_qq}, match that of the marginal distribution for $Y$ in the data, just on a narrower range.

\section{Discussion}

We introduced a new Bayesian semiparametric regression model that is designed to scale to large data applications. In doing so we make use of an interpretable model for ranks, via a Plackett-Luce copula method, and nonparametric density models for the marginals. We used a composite marginal likelihood approximation that leads to a number of advantages. It affords computationally tractability, aids in the interpretation of the model, and makes prior specification explicit on known objects.

The key to the scalability of the method is the use of the composite likelihood approximation, which splits the inference into two simpler tasks. The use of the Laplace approximation for the covariate effect and the P\'olya tree for the marginal response allow for fast posterior inference, without requiring any MCMC sampling methods. In fact, sampling methods are only used for prediction, which is by far the slowest part of the inference procedure.

Going forward, it would be interesting to see if theoretical bounds on the approximation error as a function of sample size could be derived. It may also be possible to apply results such as those found in \citet{kim2006} to provide further guarantees of asymptotic behavior such as properties of the predictive distribution. In addition, it would be interesting to explore non-linear models for the regression function $\lambda(x)$, such as those based on a random forests methodology. In fact, random forests applied to the US Census dataset (with restricted node size to enable application to this scale) gives a  highly competitive MSE to our tested models. This might be because random forests is able to capture interaction terms between covariates, which seem highly plausible {\em{a priori}} in this particular dataset. It will be interesting to incorporate such flexibility into a Bayesian nonparametric approach using Plackett-Luce regression functions.

\bibliographystyle{authordate1}
\bibliography{PLC}

\newpage
\appendix

\section{Details on simulating from the predictive}
In this section we provide additional on the simulation from the predictive distribution depending on the choice of the prior on $F_Y$.
\subsection{Empirical CDF}

It might be the case that $n$ is so large that simply using the empirical CDF is a reasonable approximation. In this case, the inversion of the cdf becomes trivial, and MCMC is only required for the $\beta$ posterior sample.

\begin{itemize}
\item Simulate $\beta^{(j)}$ from the partial posterior
\item $Z' \sim \Exp (\lambda_{\beta^{(j)}}(x'))$
\item Calculate $u^{(j)} := 1-  \frac{1}{n} \sum_{i=1}^n  e^{-Z' \lambda_{\beta^{(j)}}(x_i)}$
\item Set $Y'^{(j)} = y_{(\lceil n u^{(j)} \rceil)}$
\end{itemize}

\subsubsection{Bayesian Bootstrap}

An alternative approach might be to use a Bayesian Bootstrap on $F_Y$. This works out similarly to using the empirical CDF, but we must simulate the Dirichlet weights for each of the atoms. The sampling scheme becomes:

\begin{itemize}
\item Simulate $\beta^{(j)}$ from the partial posterior
\item Sample $Z' \sim \Exp (\lambda_{\beta^{(j)}}(x'))$
\item Calculate $u^{(j)} := 1-  \frac{1}{n} \sum_{i=1}^n  e^{-Z' \lambda_{\beta^{(j)}}(x_i)}$
\item Simulate $(W_1, W_2, \ldots, W_n) \sim \text{Dirichlet}(1,1, \ldots, 1)$
\item Set $Y'^{(j)} = y_{(\text{min}\{k : \sum_{i=1}^k W_i \geq u^{(j)}\})}$
\end{itemize}

\subsubsection{P\'olya Trees}

Under our composite likelihood scheme, the posterior for $F_Y$ is also a P\'olya tree, due to conjugacy of the P\'olya tree prior. Simulation then proceeds as follows:

\begin{itemize}
\item Simulate $\beta^{(j)}$ from the partial posterior
\item $Z' \sim \Exp(\lambda_{\beta^{(j)}}(x'))$
\item Calculate $u^{(j)} := 1-  \frac{1}{n} \sum_{i=1}^n  e^{-Z' \lambda_{\beta^{(j)}}(x_i)}$
\end{itemize}

Then all we need to calculate is $F_Y^{-1 (j)}(u^{(j)})$. P\'olya trees make this easy. A sample from a P\'olya tree distribution is a random probability measure, which is constructed by assigning random masses to each branch in a partition tree of the space. So, given the first partition point in the tree, we can simply generate the random mass either side of this point, and trivially deduce which branch $F_Y^{-1 (j)}(u^{(j)})$ lies in. We repeat this process down the tree until we reach the truncation point often used when using P\'olya trees.

Formally, given a P\'olya tree truncated at level $M$, let $a_0 = 0, B_0 = 1, \epsilon_0 = \emptyset$ and for k from 1 to M:

\begin{itemize}
\item $\theta_{\epsilon_k 0} \sim \text{Beta}(\alpha_{\epsilon_k 0 }, \alpha_{\epsilon_k  1})$
\item if $u \in [a_k, \theta_k (b_{k-1} - a_{k-1}) + a_{k-1}]$
\begin{itemize}
\item $\epsilon_k = \epsilon_{k-1} 0$
\item $a_k = a_{k-1}$
\item $b_k = a_{k-1} + \theta_k (b_{k-1} - a_{k-1})$
\end{itemize}
\item Otherwise
\begin{itemize}
\item $\epsilon_k = \epsilon_{k-1} 1$
\item $a_k = a_{k-1} + \theta_k (b_{k-1} - a_{k-1})$
\item $b_k = a_{k-1}$
\end{itemize}
\end{itemize}

\subsubsection{Dirichlet Process Mixture models}

The difficulty here becomes the inversion of $F_Y$, since this has no closed form. A simple Monte Carlo approximation can be used to approximate this inversion for each posterior sample.

\begin{itemize}
\item Simulate $\beta^{(j)}$ from the partial posterior
\item Sample $Z' \sim \Exp (\lambda_{\beta^{(j)}}(x'))$
\item Simulate $F_Y^{(j)}$ from the partial posterior
\item Simulate $Y'^{(j)}_k \sim F_Y^{(j)}$ for $k=1,\ldots, N$
\item Calculate $u^{(j)} := 1-  \frac{1}{n} \sum_{i=1}^n  e^{-Z' \lambda_{\beta^{(j)}}(x_i)}$
\item Set $Y'^{(j)} =  Y'^{(j)}_{(\lceil N u^{(j)} \rceil)}$
\end{itemize}

%
%

\end{document}